\documentclass[11pt]{article}

\usepackage{amsmath}
\usepackage{amssymb}
\usepackage{amsfonts}
\usepackage{amsthm}
\usepackage{mathtools}
\usepackage{accents}

\usepackage{dsfont}

\usepackage
{xcolor}
\colorlet{MyBlue}{blue!90!black}
\colorlet{MyGreen}{green!90!black}

\usepackage[font=small,labelfont=bf]{caption}
\usepackage{subfigure}
\usepackage{tikz}
\usetikzlibrary{calc,patterns}
\usetikzlibrary{arrows,shapes,positioning}

\usepackage{acronym}
\usepackage{booktabs}       
\usepackage{enumitem}
\usepackage{latexsym}
\usepackage{wasysym}
\usepackage{xspace}
\usepackage{fullpage}
\usepackage[onehalfspacing]{setspace}
\usepackage{longtable}

\usepackage[authoryear,comma]{natbib}

\usepackage{hyperref}
\hypersetup{
colorlinks=true,
linktocpage=true,
pdfstartview=FitH,
breaklinks=true,
pdfpagemode=UseNone,
pageanchor=true,
pdfpagemode=UseOutlines,
plainpages=false,
bookmarksnumbered,
bookmarksopen=false,
bookmarksopenlevel=1,
hypertexnames=true,
pdfhighlight=/O,
urlcolor=black, linkcolor=black, citecolor=black, 
pdftitle={},
pdfauthor={},
pdfsubject={},
pdfkeywords={},
pdfcreator={pdfLaTeX},
pdfproducer={LaTeX with hyperref}
}
\usepackage{appendix}

\numberwithin{equation}{section}  
\usepackage[sort&compress,capitalize,nameinlink]{cleveref}
\crefname{app}{Appendix}{Appendices}

\crefrangeformat{equation}{\upshape(#3#1#4)\textendash(#5#2#6)}

\usepackage[textwidth=30mm]{todonotes}
\newcommand{\debug}[1]{#1}

\usepackage{soul}

\theoremstyle{plain}
\newtheorem{theorem}{Theorem}
\newtheorem{corollary}{Corollary}
\newtheorem{lemma}{Lemma}
\newtheorem{proposition}{Proposition}

\newtheorem{observation}{Observation}

\theoremstyle{definition}
\newtheorem{definition}[theorem]{Definition}

\theoremstyle{remark}

\newtheorem{example}{Example}

\DeclarePairedDelimiter{\braces}{\{}{\}}
\DeclarePairedDelimiter{\bracks}{[}{]}
\DeclarePairedDelimiter{\parens}{(}{)}
\DeclarePairedDelimiter{\abs}{\lvert}{\rvert}

\DeclarePairedDelimiterX{\braket}[2]{\langle}{\rangle}{#1,#2}

\DeclarePairedDelimiterX{\inner}[2]{\langle}{\rangle}{#1,#2}
\DeclarePairedDelimiterX{\setdef}[2]{\{}{\}}{#1:#2}

\DeclarePairedDelimiterXPP{\probof}[1]{\Prob}{(}{)}{}{%

#1}

\DeclarePairedDelimiterXPP{\exof}[1]{\Expect}{[}{]}{}{%

#1}

\newcommand{\dirac}{\debug \delta}
\newcommand{\borel}{\mathcal{\debug B}}
\newcommand{\borelset}{\debug B}

\newcommand{\statet}{\debug \theta}

\newcommand{\states}{\debug \Theta}

\newcommand{\prior}{\debug p}

\newcommand{\simplex}{\debug \Delta}

\newcommand{\wstar}{\mathsf{\debug{w^{*}}}}
\newcommand{\wstarto}{\xrightarrow{\wstar}}

\DeclareMathOperator{\Expect}{\mathsf{\debug{E}}}
\DeclareMathOperator{\Prob}{\mathsf{\debug{P}}}
\DeclareMathOperator{\Var}{\mathsf{\debug{Var}}}

\newcommand{\act}{\debug a}
\newcommand{\actalt}{\debug b}

\newcommand{\actprof}{\boldsymbol{\debug \act}}

\newcommand{\actions}{\mathcal{\debug A}}
\newcommand{\subactions}{\mathcal{\debug B}}

\newcommand{\cost}{\debug c}
\newcommand{\costprof}{\boldsymbol{\cost}}
\newcommand{\Cost}{\debug C}
\newcommand{\Costprof}{\boldsymbol{\Cost}}

\newcommand{\outcome}{\debug \mu}
\newcommand{\outcomealt}{\debug \nu}
\newcommand{\game}{\debug \Gamma}

\newcommand{\nplayers}{\debug n}

\newcommand{\play}{\debug i}

\newcommand{\players}{\bracks{\nplayers}}

\newcommand{\potential}{\debug \Phi}

\newcommand{\signal}{\debug s}

\newcommand{\types}{\mathcal{\debug T}}

\newcommand{\weight}{\debug w}
\newcommand{\weightprof}{\boldsymbol{\weight}}

\newcommand{\eq}[1]{#1^{\ast}}

\newcommand{\DC}{f}

\DeclareMathOperator{\WE}{\mathsf{\debug{WE}}}
\DeclareMathOperator{\CWE}{\mathsf{\debug{CWE}}}
\DeclareMathOperator{\CCWE}{\mathsf{\debug{CCWE}}}

\newcommand{\edges}{\mathcal{\debug E}}

\newcommand{\edge}{\debug e}

\newcommand{\flow}{\debug y}
\newcommand{\Flow}{\debug Y}
\newcommand{\flows}{\mathcal{\Flow}}

\newcommand{\flowprof}{\boldsymbol{\flow}}

\newcommand{\flowz}{\debug z}

\newcommand{\flowprofz}{\boldsymbol{\flowz}}
\newcommand{\floww}{\debug w}
\newcommand{\flowprofw}{\boldsymbol{\floww}}

\newcommand{\load}{\debug x}

\newcommand{\lagrange}{\mathcal{\debug L}}
\newcommand{\multipl}{\debug \lambda}
\newcommand{\multiplprof}{\boldsymbol{\multipl}}
\newcommand{\primal}{\textsf{P}}

\newcommand{\reals}{\mathbb{R}}
\newcommand{\naturals}{\mathbb{N}}

\newcommand{\argdot}{\,\cdot\,}
\newcommand{\diff}{\ \textup{d}}

\newcommand{\ie}{i.e.,\ }
\newcommand{\eg}{e.g.,\ }
\newcommand{\iid}{i.i.d.\ }

\DeclareMathOperator*{\argmin}{arg\,min}

\DeclareMathOperator{\co}{co}

\DeclareMathOperator{\ind}{\mathds{1}}

\DeclareMathOperator{\supp}{supp}

\newcommand{\bce}{\debug \beta}

\newcommand{\pop}{\debug k}
\newcommand{\pops}{\mathcal{\debug K}}
\newcommand{\sizepop}{\debug \gamma}
\newcommand{\sizepopprof}{\boldsymbol{\sizepop}}
\newcommand{\prtype}{\debug \pi}

\newcommand{\spaceX}{\mathcal{\debug X}}
\newcommand{\setzc}{\mathcal{\debug Z}}
\newcommand{\closure}{\mathcal{\debug C}}
\newcommand{\nappr}{\debug N}
\newcommand{\error}{\debug \eta}

\newcommand{\modcont}{\debug \omega}

\newcommand{\recc}{\debug Z}
\newcommand{\reccprof}{\boldsymbol{\recc}}

\newcommand{\spaceJ}{\mathcal{\debug J}}
\newcommand{\elemj}{\debug j}

\newacro{AG}{anonymous game}
\newacro{ACG}{atomic congestion game}
\newacro{ACGSD}{atomic congestion game with stochastic demand}
\newacro{ANG}{nonatomic game}

\newacro{IIANG}{incomplete information anonymous nonatomic game}
\newacro{IIAG}{incomplete information anonymous game}

\newacro{BANG}{Bayesian  nonatomic game}

\newacro{PoA}{price of anarchy}
\newacro{PoS}{price of stability}
\newacro{TC}{total cost}
\newacro{TEC}{total expected cost}
\newacro{SO}{social optimum}
\newacro{SOC}{socially optimum cost}
\newacro{DC}{designer cost}

\newacro{NE}{Nash equilibrium}
\newacroplural{NE}[NE]{Nash equilibria}
\newacro{CE}{correlated equilibrium}
\newacroplural{CE}[CE]{correlated equilibria}
\newacro{CCE}{coarse correlated equilibrium}
\newacroplural{CCE}[CCE]{coarse correlated equilibria}
\newacro{PNE}{pure Nash equilibrium}
\newacroplural{PNE}[PNE]{pure Nash equilibria}

\newacro{BNE}{Bayes Nash equilibrium}
\newacroplural{BNE}[BNE]{Bayes Nash equilibria}
\newacro{BCE}{Bayes correlated equilibrium}
\newacroplural{BCE}[BCE]{Bayes correlated equilibria}
\newacro{BCCE}{Bayes coarse correlated equilibrium}
\newacroplural{BCCE}[BCCE]{Bayes coarse correlated equilibria}

\newacro{ENE}[$\varepsilon$-NE]{$\varepsilon$-Nash equilibrium}
\newacroplural{ENE}[$\varepsilon$-NE]{$\varepsilon$-Nash equilibria}
\newacro{ECE}[$\varepsilon$-CE]{$\varepsilon$-correlated equilibrium}
\newacroplural{ECE}[$\varepsilon$-CE]{$\varepsilon$-correlated equilibria}
\newacro{CCE}{coarse correlated equilibrium}
\newacroplural{CCE}[CCE]{coarse correlated equilibria}
\newacro{PNE}{pure Nash equilibrium}
\newacroplural{PNE}[PNE]{pure Nash equilibria}

\newacro{EBNE}[$\varepsilon$-BNE]{$\varepsilon$-Bayes Nash equilibrium}
\newacroplural{EBNE}[$\varepsilon$-BNE]{$\varepsilon$-Bayes Nash equilibria}
\newacro{EBCE}[$\varepsilon$-BCE]{$\varepsilon$-Bayes correlated equilibrium}
\newacroplural{EBCE}[$\varepsilon$-BCE]{$\varepsilon$-Bayes correlated equilibria}
\newacro{BCCE}{Bayes coarse correlated equilibrium}
\newacroplural{BCCE}[BCCE]{Bayes coarse correlated equilibria}

\newacro{CCWE}{coarse correlated Wardrop equilibrium}
\newacroplural{CCWE}[CCWE]{coarse correlated Wardrop equilibria}

\newacro{CBCWE}{coarse Bayes correlated Wardrop equilibrium}
\newacroplural{CBCWE}[CBCWE]{coarse Bayes correlated Wardrop equilibria}

\newacro{SBCWE}{simple Bayes correlated Wardrop equilibrium}
\newacroplural{SBCWE}[SBCWE]{simple Bayes correlated Wardrop equilibria}

\newacro{WE}{Wardrop equilibrium}
\newacroplural{WE}[WE]{Wardrop equilibria}

\newacro{BWE}{Bayesian Wardrop equilibrium}
\newacroplural{BWE}[BWE]{Bayesian Wardrop equilibria}

\newacro{CWE}{correlated Wardrop equilibrium}
\newacroplural{CWE}[CWE]{correlated Wardrop equilibria}

\newacro{BDWE}[BDWE]{Bayes deterministic Wardrop equilibrium}
\newacroplural{BDWE}[BDWE]{Bayes deterministic Wardrop equilibria}

\newacro{BCWE}{Bayes correlated Wardrop equilibrium}
\newacroplural{BCWE}[BCWE]{Bayes correlated Wardrop equilibria}

\newacro{KKT}{Karush\textendash Kuhn\textendash Tucker}
\newacro{OD}[OD]{origin-destination}
\newacro{BPR}{Bureau of Public Roads}

\newacro{SP}{series-parallel}

\begin{document}


\title
{Correlated Equilibria in Large Anonymous Bayesian Games}

\author{Frederic Koessler%
\footnote{HEC Paris and GREGHEC-CNRS, 78351 Jouy-en-Josas, France. \textsf{koessler@hec.fr}}
\and
Marco Scarsini%
\footnote{Luiss University, 00197 Rome, Italy. \textsf{marco.scarsini@luiss.it}}
\and
Tristan Tomala%
\footnote{HEC Paris and GREGHEC, 78351 Jouy-en-Josas, France. \textsf{tomala@hec.fr}}
}

\maketitle

\begin{abstract}
We consider multi-population Bayesian games with a large number of players.  Each player aims at minimizing a cost function that depends on this player's own action, the distribution of players' actions in all populations, and an unknown state parameter.  We study the  nonatomic limit versions of these games and introduce the concept of  Bayes correlated Wardrop equilibrium, which extends the concept of  Bayes correlated equilibrium to  nonatomic games. We prove that Bayes correlated Wardrop equilibria are limits of action flows induced by  Bayes correlated equilibria of the game with a large finite set of small players. For nonatomic games with complete information admitting a convex potential, we prove that the set of correlated and of coarse correlated Wardrop equilibria coincide with the set of probability distributions over Wardrop equilibria, and that all equilibrium outcomes have the same costs. We get the following consequences. First, all flow distributions of  (coarse) correlated equilibria  in  convex potential games with finitely many players converge to Wardrop equilibria when the weight of each player tends to zero.  Second, for any sequence of flows satisfying a no-regret property, its empirical distribution converges to the set of distributions over Wardrop equilibria and the average cost converges to the unique Wardrop cost.

\medskip \noindent \textsc{Keywords}: Bayes correlated equilibrium, coarse correlated equilibrium, congestion games, no-regret, nonatomic games, potential games, selfish routing, Wardrop equilibrium.

\end{abstract}

\clearpage


\section{Introduction}
\label{se:intro}

In many interactive decision situations with a  large number of participants,   single agents are impacted only by their own choice, the distribution of actions in the population, and the state of the world. 
For instance, the time  a morning commuter spends on the road to go from home to office depends on the chosen route, the number of commuters on the various roads, and the presence of accidents; 
the benefit of adopting a new technological standard or subscribing to a social network depends on the quality of the standard or network, and on the proportions of adopters  or subscribers; 
carbon emissions due to traffic depend on the proportions of people traveling by plane, by car or by train.
Such situations can be represented by anonymous games where the utility that a player enjoys, or the cost that a player incurs depends on this player's action and on the distribution of other players' actions. 
A widely studied class of games of this type is given by  congestion games, where players use resources whose cost depends on the number of users. 
Prominent applications are routing games, where players travel through a network of roads with the objective of reaching destination as  fast as possible.  
When the number of players is large and each individual has a small impact on the overall distribution, these games can be approximated by nonatomic games, which are often more tractable  \citep[see, \eg][]{Rou:AGT2007}.

In this paper, we consider a general model of Bayesian nonatomic games with multiple populations, where the action sets and cost functions are population specific.
Each agent incurs a cost depending on the individual action, on the action distribution in each population, and on an a priori unknown state parameter. 
A prominent instance are Bayesian routing games and, in line with the literature on this topic, we take the convention that players minimize cost functions, which  simply are negatives of utility functions. 
Yet, our class of games is much larger than routing games or congestion games and encompasses  population games, as described in \citet{San:MIT2010}. These include some oligopoly games and random matching games. 
We extend this class of games to the Bayesian setting. 

We study two main solution concepts, \acfi{CWE}\acused{CWE} and \acfi{BCWE}\acused{BCWE}, that allow players' actions to be correlated which one another and with the state of the world. 
They are the analogs of \acl{CE} \citep{Aum:JME1974} and \acl{BCE} \citep{BerMor:TE2016}. The concepts of \acl{CE} and \acl{BCE} in \citet{Aum:JME1974,Aum:E1987} and \citet{BerMor:TE2016} are justified by considering all possible  equilibrium outcomes for all possible information structures, in the context of games with finitely many players. 
This approach immediately faces the  daunting task of describing all information structures for a continuum of players. 
To circumvent this problem, we adopt the point of view of action flows and Wardrop equilibria \citep[called Nash equilibria of population games in][]{San:MIT2010} where  strategy profiles that map players to actions are not explicitly defined. Instead, we consider the distributions of action flows in each population that result from equilibrium behavior. Hence, our solution concepts are expressed in terms of probability distributions over action flows in each population (conditional on the state), instead of probability distributions over action profiles. 
We also introduce the concept of \emph{\acl{BDWE}}, which corresponds to a \ac{BCWE} with deterministic flows: it assigns a flow over actions to each state of the world. Under complete information, a \acl{BDWE} is a  \acl{WE}.

Our results motivate the model of nonatomic games as a limit of games with large but finite sets of players with small weights. 
We show that every converging sequence of \acl{BCE} outcomes of finite $\nplayers$-player games, converges to a \ac{BCWE} when $\nplayers$ tends to infinity and the weight of every player tends to zero (\cref{pr:convergence-A-NA}). 
Conversely, we show that for every BCWE, there exists a sequence of (approximate) \acl{BCE} outcomes that converges to that BCWE (\cref{pr:convergence-converse}). 
Furthermore, we prove that Bayes deterministic Wardrop equilibria correspond to limits of \acl{BCE} outcomes with conditionally independent signals (\cref{pr:convergence-A-NA-simple}).

In  games with complete information that admit a convex potential (i.e., the cost functions are the  partial derivatives of a convex function), we show that every coarse CWE (and therefore every CWE)  is a mixture of \aclp{WE}, and  all \aclp{WE} have the same cost profiles (\cref{pr:potential-convex-hull-equilibria}). 
As a consequence, all (coarse) CWE have the same cost profiles as well. 
This result applies in particular to congestion games with increasing resource  costs, which admit a convex potential. 
Combined with our convergence result, this result implies that all flow distributions of  correlated equilibria  in  games with finitely many players, converge to mixtures of  \aclp{WE}  (\cref{co:convergence-potential}).
These results  do not hold without a convex potential (see \cref{ex-El-Farol} and \cref{ex:trucks})  and do not extend to incomplete information (see \cref{ex-BCWEnotBDWE}).
 To the best of our knowledge, we provide the first non-trivial class of games with finite action  sets where correlated equilibria coincide with convex combinations of Nash equilibrium outcomes. 
This property does not  hold even for two-player zero-sum games  \citep[see][]{For:E1990}.
 Another consequence of \cref{pr:potential-convex-hull-equilibria} is that if a nonatomic game has a convex potential, then any accumulation point of  a sequence of flows that has no external regret,
is a mixture of \aclp{WE} (\cref{coro-no-regret}).


\subsection*{Related literature}

Congestion games with a finite number of players are introduced by \citet{Ros:IJGT1973}, their relations with potential games are studied by \citet{MonSha:GEB1996}. 
\citet{War:PICE1952} studies a strategic model of traffic where each agent has a negligible weight and introduces the principle now known as \acl{WE}. 
\citet{BecMcGWin:Yale1956} characterize \aclp{WE} as the solutions of a convex optimization program. 
The works of \citet{San:JET2001,San:MIT2010} give a general model of games with nonatomic players, beyond congestion games, and a general definition of potential games; see also \citet{HofSan:JET2009}.
Notions related to nonatomic versions of correlated and Bayes correlated equilibria have been studied under different names \citep[see, \eg][]{DiaMitRusSai:WINE2009,TavTen:mimeo2018,ZhuSav:arXiv2021}.

The relationships between \aclp{NE} and \aclp{WE} in congestion games are studied by \citet{HauMar:N1985} in the atomic splittable case, where finitely many players can split their weights among several routes, and by \citet{ComScaSchSti:MOR2023} in the atomic nonsplittable case. 
\citet{San:JET2001} studies the convergence of the potential of  finite potential games to the potential of  nonatomic potential games.

Properties of the set of \aclp{CE} in both finite and infinite games are studied by \citet{HarSch:MOR1989}. 
Various extensions of correlated equilibria to incomplete information are analyzed by \citet{Mye:JME1982}, \citet{For:TD1993,For:TD2006} and \citet{BerMor:TE2016}. 
\citet{Ney:IJGT1997}  proves that in games with a finite number of players, convex strategy sets, and a convex potential, every \acl{CE} is a mixture of \aclp{NE} (\citealp{Ney:IJGT1997}, actually considers players who maximize utilities in a game with a concave potential; for the equivalent game where players minimize costs, the potential is convex).
This result is extended under the weaker condition of diagonal convexity by  \citet{Ui:IJGT2008}.

\citet{AshMonTen:JAIR2008} introduce the mediation value---\ie the ratio between the maximal welfare in a \acl{CE} and the maximal welfare in a mixed-strategy equilibrium---and  show that the mediation value can be arbitrarily high in congestion games with finitely many players.
A result closely related to our  \cref{pr:potential-convex-hull-equilibria} is in \citet{DiaMitRusSai:WINE2009} who show  that correlation does not decrease total cost compared to Nash equilibria in congestion games with parallel edges and nonatomic players (they use a definition of correlated equilibrium which is equivalent to a \acl{CWE} with finite support). 
This result does not extend beyond congestion games, as shown by the El Farol example given in  \citet{DiaMitRusSai:WINE2009} and \citet{MitSaaSai:SAGT2013}, which we recall in \cref{ex-El-Farol}.

In our paper, we abstract away from the description of the set of players as a measurable space. 
Instead, we focus on action flows, as is common in the literature on congestion games \citep{Rou:AGT2007}, and as is done in the systematic study of population games \citep{San:MIT2010}.
There exists a huge literature on games with measurable set of players, and \citet{HarSch:MOR1989} have extended the definition of correlated equilibrium to measurable sets of players, using  finitely additive measures.
It is not difficult to check that any countably additive correlated equilibrium of \citet{HarSch:MOR1989} induces a \acl{CWE}; the proof is available upon request.
The approach of describing general information structures for arbitrary measurable sets of players inevitably faces the technical hurdle of uncountable families of independent random variables. 
We have purposefully chosen to avoid those difficulties. 
For a recent contribution on related topics, see \citet{Hel:TE2022} and references therein.


\subsection*{Organization of the paper}

\cref{se:anonymous-nonatomic} presents the multi-population model of Bayesian nonatomic games and defines the concept of Bayes correlated Wardrop equilibrium and related equilibrium notions. 
\cref{se:convergence} shows our convergence results, which prove that the concept of Bayes correlated Wardrop equilibrium is a reasonable approximation of Bayes correlated equilibria in games with large finite sets of players. 
\cref{se:complete} deals with complete-information games that admit a convex potential.
\cref{se:symbols} contains a list of symbols used throughout the paper. 


\subsection*{Notation}

For any compact set $\spaceX$, we let $\simplex(\spaceX)$ be the set of Borel probability distributions over $\spaceX$. 
The symbol $\dirac_x$ denotes the Dirac mass on $x$. 
Given probability distributions  $\{\outcome^{\nplayers}\}_{\nplayers\in\naturals},\outcome$  in $\simplex(\spaceX)$, the notation $\outcome^{\nplayers} \wstarto \outcome$ denotes weak$^{*}$ convergence of $\outcome^{\nplayers}$ to $\outcome$ \ie $\int f \diff\outcome^{\nplayers}\to\int f \diff\outcome$ for every continuous function $f \colon \spaceX\to\reals$.
For a finite set $\spaceJ$ we let $\abs{\spaceJ}$ denote its cardinality and, for $\sizepop>0$, we use the notation
\begin{equation}
\label{eq:simplex-gamma}
\simplex_{\sizepop}(\spaceJ) \coloneqq \braces*{\flowprof\in\reals^\spaceJ \colon \forall \elemj\in \spaceJ, \flow_{\elemj} \geq 0, \sum_{\elemj\in\spaceJ} \flow_{\elemj}  = \sizepop}.
\end{equation}
Therefore, in this case, we identify $\simplex(\spaceJ)$ with $\simplex_{1}(\spaceJ)$.


\section{Bayesian nonatomic games}
\label{se:anonymous-nonatomic}

We study Bayesian nonatomic games where  infinitely many players  are partitioned into finitely many populations, and each population has a finite set of available actions. 
The analog of a strategy profile in this context is a distribution of actions in each population, which we call a \emph{flow}. 
Players aim at minimizing cost functions that are population specific and depend on flows and on an unknown state parameter drawn from a finite set.
This section formally presents our model  and  solution concepts.

\begin{definition}
\label{de:BANG}    
A \emph{\acl{BANG}} $\game = \parens*{\pops, \sizepopprof, \actions, \states,\prior,\costprof}$ is given by the following elements:
\begin{enumerate}[label=(\alph*)]
\item 
A finite set $\pops$ of \emph{populations}, where each population $\pop\in\pops$ has size $\sizepop^{\pop}>0$.

\item
For each population $\pop\in\pops$, a finite set of \emph{actions} $\actions^{\pop}$, with $\actions = \times_{\pop\in\pops} \actions^{\pop}$. For each  $\pop\in\pops$, the set of  \emph{flows} is 
\begin{equation}
\label{eq:flows}  
\flows^{\pop} \coloneqq \simplex_{\sizepop^{\pop}}(\actions^{\pop}), 
\end{equation}
and $\flows=\times_{\pop\in\pops} \flows^{\pop}$ is the set of flow profiles.

\item
A finite set of \emph{states} $\states$ and a full-support \emph{probability distribution} $\prior\in\simplex(\states)$ over states.

\item
For each $(\pop,\act) \in \pops \times \actions^{\pop}$, a continuous \emph{cost function} $\cost_{\act}^{\pop}: \flows \times \states \to\reals$ (given a flow profile $\flowprof$, in state $\statet$, the cost for an individual in population $\pop$ who chooses action $\act$ is $\cost_{\act}^{\pop}(\flowprof,\statet)$).

\end{enumerate}
\end{definition}
A transition  probability $\outcome \colon \states\to \simplex(\flows)$ that associates a distribution over flows to each state is called an \emph{outcome} of the game. 

\begin{definition}
\label{de:BCWE}
 A  \acfi{BCWE}\acused{BCWE} of a Bayesian nonatomic game $\game$ is an outcome  $\outcome:\states\to\simplex(\flows)$ such that for every $\pop\in \pops$ and $\act, \actalt\in \actions^{\pop}$:
 \begin{equation}\label{eq:BCWEmutltipop} 
\sum_{\statet\in\states} \prior(\statet)\int \flow_{\act}^{\pop} \cost_{\act}^{\pop}(\flowprof,\statet)\diff\outcome(\flowprof  \mid\statet)
\leq\sum_{\statet\in\states} \prior(\statet) \int \flow_{\act}^{\pop} \cost_{\actalt}^{\pop}(\flowprof,\statet)\diff\outcome(\flowprof \mid \statet).
\end{equation}
\end{definition}

The interpretation of the definition of \ac{BCWE} is as follows. 
For each state $\statet\in\states$, a mediator who knows the state draws a flow $\flowprof$ at random according to the distribution $\outcome(\,\cdot\mid\statet)$. 
Then, for each population $\pop\in\pops$ and action $\act\in\actions^\pop$, the mediator recommends a random mass $\flow_{\act}^{\pop}$ of  players in population $\pop$ to play $\act$.
The outcome is a \ac{BCWE} if no player has an incentive to deviate from the mediator's recommendation when all the other players follow the mediator's recommendation. 
Indeed, for each population $\pop$, if we divide \cref{eq:BCWEmutltipop}  by the ex-ante expected mass $\sum_{\statet} \prior(\statet) \int\flow_{\act}^{\pop}\diff\outcome(\flowprof\mid \statet)$  of players choosing $\act$ in population $\pop$, then we see that the left-hand-side is proportional to the  expected cost of playing $\act$, conditionally of being recommended $\act$, and the right-hand-side is proportional to the  expected cost of playing $\actalt$, conditionally of being recommended $\act$. 

The definition of \ac{BCWE} has  some particular cases of interest. First, a \acfi{BDWE}\acused{BDWE} is a mapping $\flowprof(\argdot) \colon \states \to \flows$ such that for every $\pop\in \pops$ and $\act, \actalt\in \actions^{\pop}$:
 \begin{equation}
\label{eq:BDWE-def}
\sum_{\statet\in\states} \prior(\statet) \flow_{\act}^{\pop}(\statet) \cost_{\act}^{\pop}(\flowprof(\statet),\statet) 
\le
\sum_{\statet\in\states} \prior(\statet) \flow_{\act}^{\pop}(\statet) \cost_{\actalt}^{\pop}(\flowprof(\statet),\statet).
\end{equation}
Hence, a \ac{BDWE} is a \ac{BCWE} such that in each state the distribution $\outcome(\,\cdot\mid\statet)$ is a Dirac distribution $\dirac_{\flowprof(\statet)}$ on some flow  $\flowprof(\statet)$. 
The  \ac{BDWE} inequalities can be interpreted as obedience constraints when, in each state $\statet$, the mass of players receiving recommendation $\act$ in population $\pop$ is $\flow_{\act}^{\pop}(\statet)$. 
In \cref{pr:convergence-A-NA-simple}, we provide a formal link between  \ac{BDWE} and equilibria in games with large finite sets of players  who get conditionally independent signals. 

Second, under complete information (\ie $\states = \braces*{\statet}$),  the definition of \ac{BDWE} reduces to the well known definition of \acfi{WE}\acused{WE}. A \ac{WE} of the game with complete information  at state $\statet$, is a flow $\flowprof=(\flowprof^{\pop})_{\pop\in \pops}$ such that for all $\pop$, and all $\act$, $\actalt$ in  $\actions^{\pop}$
\begin{equation}
\label{eq:WE-multi}
\flow_{\act}^{\pop} \cost_{\act}^{\pop}(\flowprof,\statet)\leq \flow_{\act}^{\pop} \cost_{\actalt}^{\pop}(\flowprof,\statet).
\end{equation}
In other words, in each population, only actions with the smallest cost receive a positive flow:
\begin{equation*}
\forall \pop\in\pops,\forall\act\in\actions^{\pops},\ \flow_{\act}^{\pop}>0\implies \cost_{\act}^{\pop}(\flowprof,\statet)=\min_{\actalt} \cost_{\actalt}^{\pop}(\flowprof,\statet).
\end{equation*}
This is called a Nash equilibrium of population games in \citet{San:MIT2010}.

Existence of \acp{WE} is easily proved by a standard Kakutani fixed point argument \citep[see][theorem~2.1.1, page 24]{San:MIT2010}. 
Observe that if, for every $\statet\in\states$, $\flowprof(\statet)$ is a \ac{WE} of the nonatomic game with complete information  at $\statet$, then $\flowprof(\argdot)$ is a \ac{BDWE} of $\game$. 
Hence, for every Bayesian nonatomic game,  \acp{BDWE} and \acp{BCWE} exist.

Congestion games are important instances of our general model.
A \emph{Bayesian congestion game} is defined by the following quantities:
\begin{itemize}
\item
a finite set of \emph{resources} $\edges$;

\item
for each population $\pop\in\pops$, a set of actions  $\actions^\pop \subseteq 2^{\edges}$;

\item
for each resource $\edge\in\edges$, a cost function $\cost_{\edge} \colon \reals_{+}\times\states \to \reals_{+}$, which is continuous in the first variable.

\end{itemize}
 
For each resource $\edge\in\edges$,  the \emph{load} on resource $\edge$ induced by the flow $\flowprof$ is 
\begin{equation}
\label{eq:load-congestion}
\load_{\edge} \coloneqq \sum_\pop \sum_{\act\in \actions^\pop, \act\ni\edge} \flow_{\act}^{\pop}.
\end{equation}
When the state is $\statet$, the cost of using resource $\edge$ is $\cost_{\edge}(\load_{\edge},\statet)$ and the cost of choosing action $\act \in \actions^\pop$ in population $\pop$ is obtained additively: 
\begin{equation}
\cost_{\act}^{\pop}(\flowprof,\statet) \coloneqq \sum_{\edge\in\act} \cost_{\edge}(\load_{\edge},\statet).
\end{equation}

One of the main applications  of congestion games is traffic routing, as is reflected in the terminology. 
In a  network routing model,  the resources are the edges of an underlying oriented multigraph; each population is represented by an origin-destination pair of vertices; and  actions  for population $\pop$ are feasible paths from its origin to its destination. 
In this case, the cost functions $\cost_{\edge}$ are nondecreasing.
The class of congestion games is larger than the class of routing games. 
For instance it includes cost-sharing games, for which the cost functions $\cost_{\edge}$ are nonincreasing.

Some authors  \citep[\eg][]{Daf:TS1980,SohHayBeaJea:IEEETITS}  consider congestion games with non-separable costs where the cost of an action is an arbitrary  continuous function of loads on resources. This generalization of congestion games is also a particular case of our model.


\section{Convergence}
\label{se:convergence}

The aim of this section is to explore the relation between Bayesian games with finitely many players and \aclp{BANG}. 
In particular, we will prove that the solution concepts introduced in \cref{se:anonymous-nonatomic} correspond to limits of equilibrium outcomes in Bayesian games with finitely many players, as the number of players tends to infinity and the players' weights tend to zero.

Given a \acl{BANG} $\game$, we consider an $\nplayers$-player Bayesian game $\game^\nplayers$ with $\abs*{\pops}$ populations, where population $\pop$ has $\nplayers^{\pop}$ players and $\nplayers=\sum_{\pop} \nplayers^{\pop}$. The action set of each player in population $\pop$ is $\actions^{\pop}$.
For each $\pop$, player $\play=1,\dots, \nplayers^{\pop}$ in population $\pop$ has  a weight $\weight_{\play}^{\pop}$ with $\sum_{\play=1}^{\nplayers^{\pop}} \weight_{\play}^{\pop}=\sizepop^{\pop}$. 

Every action profile $\actprof =(\actprof^{\pop})_{\pop} \in\bigtimes_{\pop} (\actions^{\pop})^{\nplayers^{\pop}}$ induces a flow $\flowprof(\actprof)=(\flowprof^{\pop}(\actprof^{\pop}))_{\pop}=\parens*{\parens*{\flow_{\act}^{\pop}}_{\act\in\actions^{\pop}}}_{\pop\in\pops}$ defined by
\begin{equation}
\label{eq:flow}
\flow_{\act}^{\pop} =\sum_{\play = 1}^{\nplayers^{\pop}} \weight_{\play}^{\pop}\ind\braces*{\act_{\play}^{\pop}=\act},
\end{equation}
for each $\pop\in\pops$ and $\act\in \actions^{\pop}$. 
 For each action profile $\actprof$ and state $\statet$, the cost of player $\play$ in population $\pop$ who plays action $\act$ is given by $\cost^{\pop}_{\act}(\flowprof(\actprof),\statet)$.

We now recall the definition of Bayes correlated equilibrium for finite games \citep{BerMor:TE2016} within the present context.

\begin{definition}
\label{de:BCE}
A \acfi{BCE}\acused{BCE} of the $\nplayers$-player game $\game^{\nplayers}$ is a mapping $\bce^\nplayers:\states\to \simplex(\bigtimes_{\pop} (\actions^{\pop})^{\nplayers^{\pop}})$ such that $\forall {\pop}, \forall \play=1,\dots, \nplayers^{\pop}, \forall \act,\actalt\in \actions^{\pop}$, we have
\begin{multline}
\label{eq:BCE}
\sum_{\statet, \actprof_{-\play}}\prior(\statet)\bce^\nplayers(\act,\actprof_{-\play}^{\pop}, \actprof^{-\pop}\mid\statet)\cost^{\pop}_{\act}(\flowprof^{\pop}(\act,\actprof_{-\play}^{\pop}),\flowprof^{-\pop}(\actprof^{-\pop}),\statet)\\
\leq
\sum_{\statet, \actprof_{-\play}}\prior(\statet)\bce^\nplayers(\act,\actprof_{-\play}^{\pop}, \actprof^{-\pop}\mid\statet)\cost^{\pop}_{\actalt}(\flowprof^{\pop}(\actalt,\actprof_{-\play}^{\pop}),\flowprof^{-\pop}(\actprof^{-\pop}),\statet).
 \end{multline}
When all the inequalities are satisfied up to $\varepsilon>0$ the mapping $\bce^\nplayers$ is  called an \emph{$\varepsilon$-\acl{BCE}}.

Given a mapping $\bce^{\nplayers} \colon \states \to \simplex(\bigtimes_{\pop} (\actions^{\pop})^{\nplayers^{\pop}})$, the induced \emph{outcome} is  $\outcome^\nplayers : \states\to \simplex(\flows)$ defined as:
\begin{equation}
\label{eq:BCEoutcome}
\outcome^\nplayers(\flowprof\mid\statet)=\sum_{\actprof : \flowprof(\actprof)=\flowprof}\bce^\nplayers(\actprof\mid\statet).
\end{equation}
 \end{definition}

In words, there is mediator who knows the state, draws an action profile $\actprof=(\act^{\pop}_{\play},\actprof_{-\play}^{\pop},\actprof^{-\pop})$, and privately recommends action $\act_{\play}^{\pop}\in \actions^{\pop}$ to player $\play$ in population $\pop$.  
\cref{eq:BCE} is the  \ac{BCE} constraint stating that player $\play$ who is recommended $\act_{\play}^{\pop}=\act$, prefers playing $\act$ over playing any other action $\actalt\in \actions^{\pop}$.  
Similarly to \cref{eq:BCWEmutltipop}, by dividing both sides by the probability of recommending $\act$, we see that \cref{eq:BCE} compares the expected costs of playing $\act$ or $\actalt$, conditionally on being recommended to play $\act$. 
\citet{BerMor:TE2016} proved that the set  of \ac{BCE} outcomes is the set of  all  Bayesian Nash equilibrium outcomes, for  all possible information structures  over the state set  $\states$, where an information structure is family of stochastic signals $\parens*{\signal_{\play}}_{\play}$, jointly correlated with the state and such that $\signal_{\play}$ is  privately observed by  player $\play$.

The next proposition is related to  the convergence result of \citet{ComScaSchSti:MOR2023}, which  proves the convergence of \aclp{NE} of  congestion games with finitely many players to \aclp{WE} of the corresponding nonatomic congestion games. 
The argument of the proof is  that the obedience conditions of the \ac{BCE}  of the $\nplayers$-player game depend on the distribution of flows and carry over to the limit as the number of players grows.

\begin{proposition}
\label{pr:convergence-A-NA}
Consider a sequence of weights $\braces*{\weightprof(\nplayers)}$ such that $\max_{\pop,\play}\weight_{\play}^{\pop}(\nplayers)\to_\nplayers 0$, and let $\outcome^{\nplayers}$ be a \ac{BCE} outcome of the game $\game^{\nplayers}$ for each integer $\nplayers$.
Then, any weak$^{*}$ accumulation point $\outcome$ of the sequence $\braces*{\outcome^{\nplayers}}$ is a \ac{BCWE} of the \acl{BANG} $\game$.
\end{proposition}

\begin{proof}
For given $\nplayers$, consider a \ac{BCE} outcome of the game $\game^{\nplayers}$.
Let $\Prob^{\nplayers}$ denote the probability measure induced by the prior $\prior$ and the \ac{BCE} $\bce^{\nplayers}$, and let $\Expect^{\nplayers}$ be the corresponding expectation.

\cref{eq:BCE} can be rewritten as follows:
\begin{equation}
\label{eq:BCE-weight-alt}
\Expect^{\nplayers}\bracks*{\ind\braces*{\act_{\play}^{\pop}=\act}\cost_{\act}^{\pop}(\flowprof(\actprof),\statet)}
\leq 
\Expect^{\nplayers}\bracks*{\ind\braces*{\act_{\play}^{\pop}=\act}\cost_{\actalt}^{\pop}\parens*{\flowprof^{\pop}(\actprof^{\pop}) 
+ \weight_{\play}^{\pop}(\nplayers)(\dirac^{\pop}_{\actalt}-\dirac_{\act}^{\pop}),\flowprof^{-{\pop}}( \actprof^{-{\pop}}),\statet}},
\end{equation}
where $\dirac_{\act}^{\pop}$ is the flow on $\actions^{\pop}$ such that $\flow_{\act}^{\pop}=1$. 
Multiplying by $\weight_{\play}^{\pop}(\nplayers)$ and summing over $\play=1,\dots, \nplayers^{\pop}$, we get
\begin{multline}
\label{eq:BCE-weight-alt-2}
\Expect^{\nplayers}\bracks*{\sum_{\play}\weight_{\play}^{\pop}(\nplayers)\ind\braces*{\act_{\play}^{\pop}=\act}\cost_{\act}^{\pop}(\flowprof(\actprof),\statet)}
\\
\leq \Expect^{\nplayers}\bracks*{\sum_{\play}\weight_{\play}^{\pop}(\nplayers)\ind\braces*{\act_{\play}^{\pop}=\act}\cost_{\actalt}^{\pop} \parens*{\flowprof^{\pop}(\actprof^{\pop})
+\weight_{\play}^{\pop}(\nplayers)(\dirac^{\pop}_{\actalt}-\dirac_{\act}^{\pop}),\flowprof^{-{\pop}}( \actprof^{-{\pop}}),\statet}}.
\end{multline}
The l.h.s.\ of \cref{eq:BCE-weight-alt-2} is
\begin{equation}
\label{eq:lhs}
\begin{split}
&\Expect^{\nplayers}\bracks*{\sum_{\play} \weight_{\play}^{\pop}(\nplayers)\ind\braces*{\act_{\play}^{\pop}=\act}\cost_{\act}^{\pop}(\flowprof(\actprof),\statet)} 
\\
&\qquad=\sum_{\statet,\flowprof}\Prob^{\nplayers}\parens*{\statet,\flowprof(\actprof)=\flowprof}\Expect^{\nplayers}\bracks*{\sum_{\play}\weight_{\play}^{\pop}(\nplayers)\ind\braces*{\act_{\play}^{\pop}=\act} \cost_{\act}^{\pop}(\flowprof(\actprof),\statet) \mid \flowprof(\actprof)=\flowprof,\statet}\\
&\qquad=\sum_{\statet,\flowprof}\Prob^{\nplayers}(\statet,\flowprof(\actprof)=\flowprof)\Expect^{\nplayers}\bracks*{\sum_{\play}\weight_{\play}^{\pop}(\nplayers)\ind\braces*{\act_{\play}^{\pop}=\act}\mid \flowprof(\actprof)=\flowprof} \cost_{\act}^{\pop}(\flowprof,\statet)\\
&\qquad=\sum_{\statet,\flowprof}\Prob^{\nplayers}(\statet,\flowprof(\actprof)=\flowprof)\flow_{\act}^{\pop}\cost_{\act}^{\pop}(\flowprof,\statet)\\
&\qquad=\sum_{\statet,\flowprof} \prior(\statet)\outcome^{\nplayers}(\flowprof\mid\statet) \flow_{\act}^{\pop} \cost_{\act}^{\pop}(\flowprof,\statet)\\
&\qquad=\sum_{\statet} \prior(\statet)\int \flow_{\act}^{\pop} \cost_{\act}^{\pop}(\flowprof,\statet) \diff\outcome^{\nplayers}(\flowprof\mid\statet),
\end{split}
\end{equation}
where $\outcome^{\nplayers}(\,\cdot\mid\statet)\in\simplex(\flows)$ is the (marginal) distribution of the flow $\flowprof(\actprof)$ induced by $\bce^{\nplayers}(\,\cdot\mid\statet)$. 
Consider a sequence $\{\bce^{\nplayers}\}$ of \acp{BCE} of the $\nplayers$-player game such that  for each $\statet$,  $\outcome^\nplayers(\,\cdot\mid\statet)$ weak$^{*}$ converges  to some $\outcome(\,\cdot\mid\statet)$ in $\flows$ (take a subsequence if needed) and suppose that $\max_{\pop,\play}\weight_{\play}^{\pop}(\nplayers)\to_\nplayers 0$. 
Then, the expression in \eqref{eq:lhs} tends to $\sum_{\statet} \prior(\statet)\int \flow_{\act}^{\pop} \cost_{\act}^{\pop}(\flowprof,\statet)\diff\outcome(\flowprof\mid\statet)$. 

The r.h.s.\ of \cref{eq:BCE-weight-alt-2} is
\begin{equation}
\label{eq:rhs}
\begin{split}
&\Expect^{\nplayers}\bracks*{\sum_{\play}\weight_{\play}^{\pop}(\nplayers)\ind\braces*{\act_{\play}^{\pop}=\act} \cost_{\actalt}^{\pop}(\flowprof(\actprof),\statet)}+\\
&\qquad  \Expect^{\nplayers}\bracks*{\sum_{\play}\weight_{\play}^{\pop}(\nplayers)\ind\braces*{\act_{\play}^{\pop}=\act}\parens*{\cost_{\actalt}^{\pop} \parens*{\flowprof^{\pop}(\actprof)+\weight_{\play}^{\pop}(\nplayers)(\dirac^{\pop}_{\actalt}-\dirac_{\act}^{\pop}),\flowprof^{-{\pop}}(\actprof),\statet} - \cost_{\actalt}^{\pop}(\flowprof(\actprof)),\statet}}\\
&\qquad=\sum_{\statet} \prior(\statet)\int \flowprof_{\act}^{\pop} \cost_{\actalt}^{\pop}(\flowprof,\statet) \diff\outcome^{\nplayers}(\flowprof\mid\statet)+\text{error term}.
\end{split}
\end{equation}
The first term in \cref{eq:rhs} tends to $\sum_{\statet} \prior(\statet)\int \flow_{\act}^{\pop} \cost_{\actalt}^{\pop}(\flowprof,\statet) \diff\outcome(\flowprof\mid\statet)$. 
For the second term, note that the finite family of  functions $\{\flowprof\mapsto \cost_{\actalt}^{\pop}(\flowprof,\statet)\}_{{\pop},\statet,\actalt}$ is uniformly equicontinuous on the compact $\flows$; that is,  $\forall \varepsilon>0$, $\exists \bar\weight$, s.t. $\forall \weight\le \bar\weight$, $\forall \pop,\statet,\act,\actalt$, $\forall \flowprof\in \flows$, we have
$\abs*{\cost_{\actalt}^{\pop}\parens*{\flowprof^{\pop}+ \weight(\dirac^{\pop}_{\actalt}-\dirac_{\act}^{\pop}),\flowprof^{-{\pop}},\statet}- \cost^{\pop}_{\actalt}(\flowprof,\statet)}\leq\varepsilon$. 
Thus for all $\varepsilon>0$, there exists $\bar\nplayers$ such that $\forall \nplayers\ge \bar\nplayers$, $\forall \pop,\play$, we have $\weight_{\play}^{\pop}(\nplayers)\leq \bar\weight$; hence,  $\abs*{\text{error term}}\leq\varepsilon$ for all $\nplayers\ge \bar \nplayers$. 
Therefore, the r.h.s.\ of \cref{eq:BCE-weight-alt-2} tends to $\sum_{\statet} \prior(\statet)\int \flow_{\act}^{\pop} \cost_{\actalt}^{\pop}(\flowprof,\statet) \diff\outcome(\flowprof\mid\statet)$, which proves that $\outcome$ is a \ac{BCWE}. 
\end{proof}

The next proposition shows that a \ac{BDWE} of a nonatomic game is related to a \ac{BCE} with conditionally independent signals in $\nplayers$-player games. 
Conditionally independent signals are common in the literature  on games with incomplete information such as global games or models of social learning.
A \ac{BCE} $\bce^{\nplayers}$ of the $\nplayers$-player game has \emph{the conditional independence property} if for all $\actprof$ and $\statet$, $\bce^{\nplayers}(\actprof \mid \statet) = \prod_{\pop} \prod_{\play = 1}^{\nplayers^{\pop}} \bce_{\play,\pop}^{\nplayers}(\act_{\play} \mid \statet)$  for some $\bce_{\play,\pop}^{\nplayers} \colon \states \to \simplex(\actions^{\pop})$, $\pop\in \pops$, $\play = 1,\ldots,\nplayers^{\pop}$.
That is, recommended actions are independent, conditional on the state. 
A \ac{BCE} with the conditional independence property is a Bayesian Nash equilibrium induced by conditionally independent signals.
Compared with the proof of \cref{pr:convergence-A-NA}, the additional ingredient in the proof of \cref{pr:convergence-A-NA-simple} is that, from the weak law of large numbers, conditional independence implies that distributions of flows  converge to their expectation. 
Thus, the limits of expected flows define a \ac{BDWE}.

\begin{proposition}
\label{pr:convergence-A-NA-simple}
Consider a sequence of weights $\braces*{\weight(\nplayers)}$ such that $\max_{\pop,\play}\weight_{\play}^{\pop}(\nplayers)\to_\nplayers 0$, and let  $\outcome^{\nplayers}$ be a \ac{BCE} outcome of the game $\game^{\nplayers}$ with the conditional independence property for each integer $\nplayers$.
Then, any weak$^{*}$ accumulation point $\outcome$ of the sequence $\braces*{\outcome^{\nplayers}}$ is a \ac{BDWE} of the  \acl{BANG}   
 $\game$. 
\end{proposition}

\begin{proof}
To prove this convergence result we use \cref{pr:convergence-A-NA} and show that  any  weak$^{*}$ accumulation point $\outcome$ of $(\outcome^{\nplayers}(\argdot\mid\statet))_{\statet}$   is a mapping $\flowprof \colon \states\to \flows$. 
  In other words, the distribution  of flows $\outcome(\argdot\mid\statet)$ in state $\statet$ is a Dirac mass on some flow $\flowprof(\statet)$.

Consider a sequence of  \acp{BCE} $\bce^{\nplayers}$  with outcome $\outcome^{\nplayers}$ that has the  conditional independence property for each $\nplayers$ and, up to extracting a subsequence,  assume  that  $\outcome^{\nplayers}(\argdot\mid\statet)$  converges to $\outcome(\argdot\mid\statet)$ for each $\statet\in\states$.  
For each $\pop$ and  action $\act\in \actions^{\pop}$, let $\Flow_{\act}^{\pop}(\nplayers) \coloneqq \sum_{\play=1}^{\nplayers^{\pop}}\weight_{\play}^{\pop}(\nplayers)\ind\{\act_{\play}^{\pop}=\act\}$ denote the random flow of population $\pop$ on action $\act$. 
We have $\Expect^{\nplayers}\bracks*{\Flow_{\act}^{\pop}(\nplayers)\mid\statet} = \int \flow_{\act}^{\pop} \diff\outcome^{\nplayers}(\flow \mid \statet)\to \int \flow_{\act}^{\pop} \diff\outcome(\flow\mid\statet) =: \flow_{\act}^{\pop}(\statet)$. 
To prove the result, it is enough to show that conditionally on $\statet$, the variance of $\Flow_{\act}^{\pop}(\nplayers)$ tends to 0. 
\begin{equation}
\label{eq:variance}
\Var^{\nplayers}_{\statet}\bracks*{\Flow_{\act}^{\pop}(\nplayers)} 
\leq\frac{\sum_{\play}(\weight_{\play}^{\pop}(\nplayers))^{2}}{4}
\leq\frac{(\max_{\play}\weight_{\play}^{\pop}(\nplayers))\sum_{\play}\weight_{\play}^{\pop}(\nplayers)}{4}
=\frac{\max_{\play}\weight_{\play}^{\pop}(\nplayers) \sizepop^{\pop}}{4}
\to 0.
\end{equation}
 Thus, conditionally on $\statet$, the random variable $\Flow_{\act}^{\pop}(\nplayers)$ converges in distribution to $\flow_{\act}^{\pop}(\statet)$, which means that $\outcome^{\nplayers}(\argdot\mid\statet)$ weak$^{*}$ converges  to $\dirac_{\flow(\statet)}$.
 \end{proof}

\cref{pr:convergence-A-NA-simple} implies the following corollary.  
\begin{corollary}\label{cor:convergence-WE}
Let $\Gamma$ be a game with complete information, \ie let $|\states|=1$. 
Consider a sequence of weights $\braces*{\weight(\nplayers)}$ such that $\max_{\pop,\play}\weight_{\play}^{\pop}(\nplayers)\to_\nplayers 0$, and let  $\outcome^{\nplayers}$ be a mixed Nash equilibrium outcome of the game $\game^{\nplayers}$ for each integer $\nplayers$.
Then, any weak$^{*}$ accumulation point $\outcome$ of the sequence $\braces*{\outcome^{\nplayers}}$ is a $\ac{WE}$ of $\game$. 
\end{corollary}

This corollary extends the convergence result of \citet{ComScaSchSti:MOR2023} from congestion games to our general class of nonatomic games with continuous cost functions.
The proof follows immediately from Proposition \ref{pr:convergence-A-NA-simple}, by noticing that  in a game with complete information, a \acp{BCE} is simply a correlated equilibrium, and that the conditional independence property  reduces its outcome to a Nash equilibrium in mixed strategies.

The next proposition complements  \cref{pr:convergence-A-NA,pr:convergence-A-NA-simple} by showing  converse results for approximate equilibria. 
That is, any $ \ac{BCWE}$ is a limit of approximate \ac{BCE} of  games with $\nplayers$ players. 
The structure of the proof is as follows. 
First, we approximate a \ac{BCWE} $\outcome$ with a finite-support outcome whose probabilities are rational numbers having common denominator $\nplayers$. 
Then, we consider a mediator who draws a flow from $\outcome$ and assigns actions to subsets of players drawn according to the above rational probabilities. 
If the approximation is fine enough, this yields an approximate \ac{BCE}. 
The analog of this result is proved for \ac{BDWE} by relying on the law of large numbers when the mediator sends \iid recommendations to players. 
In the limit, the mass of players who are recommended a given action coincides with the flow on this action given by the \ac{BDWE}. 

\begin{proposition}
\label{pr:convergence-converse} 
\begin{enumerate}[label=\rm(\alph*)]
\item 
\label{it:pr:convergence-converse-a}
Let $\outcome$ be a \ac{BCWE} of the \acl{BANG}   
$\game$. 
Then there exist a sequence of weights $\braces*{\weight(\nplayers)}$ such that $\max_{\pop,\play}\weight_{\play}^{\pop}(\nplayers)\to_\nplayers 0$, a sequence $\varepsilon^{\nplayers} \searrow 0$, and a sequence of $\varepsilon^{\nplayers}$-\acp{BCE} outcomes $\outcome^{\nplayers}$  of $\game^{\nplayers}$   such that  $\braces*{\outcome^{\nplayers}}$ weak$^{*}$ converges  to $\outcome$ as $\nplayers$ tends to $\infty$.

\item
\label{it:pr:convergence-converse-b}
Let $\outcome$ be a \ac{BDWE} of the  \acl{BANG}   
  $\game$. 
Then there exist a sequence of weights $\braces*{\weight(\nplayers)}$ such that $\max_{\pop,\play}\weight_{\play}^{\pop}(\nplayers)\to_\nplayers 0$, a sequence $\varepsilon^{\nplayers} \searrow 0$, and a sequence of $\varepsilon^{\nplayers}$-\acp{BCE} outcomes $\outcome^{\nplayers}$  of $\game^{\nplayers}$ with the conditional independence property such that $\braces*{\outcome^{\nplayers}}$ weak$^{*}$ converges  to $\outcome$ as $\nplayers$ tends to $\infty$.
\end{enumerate}
\end{proposition}

To prove \cref{pr:convergence-converse}, we first provide two preliminary lemmas. Let $\DC \colon \flows \times \states \to \reals$ be a  continuous function and consider the following minimization program over the set of \ac{BCWE}:
\begin{equation}
\label{eq:designer-program}
\tag{\primal}
\begin{split}
&\min_{\outcome\in\simplex(\flows)^ \states} \sum_{\statet\in\states} \prior(\statet) \int\DC(\flowprof,\statet) \diff\outcome(\flowprof\mid\statet)
\\
&\ \text{s.t.} 
\sum_{\statet\in\states} \prior(\statet) \int \flow^{\pop}_{\act}\cost^{\pop}_{\act}(\flowprof,\statet) \diff\outcome(\flowprof\mid\statet) \le
\sum_{\statet\in\states} \prior(\statet) \int \flow^{\pop}_{\act}\cost^{\pop}_{\actalt}(\flowprof,\statet) \diff\outcome(\flowprof\mid\statet),
\quad \forall \pop\in\pops ,\forall \act,\actalt\in \actions^{\pop}.
\end{split}
\end{equation}

The next lemma proves the existence of a solution with finite support, whose cardinality is upper bounded. The proof follows from a use of Caratheordory's theorem which is common in the theory-of-moment problems \citep[see, e.g.,][]{Las:MPB2008}.
The argument is based on the observation that the objective function and the constraints  depend only on the expected values of the following functions of flows: 
 \begin{equation*}
 \sum_{\statet\in\states} \prior(\statet)\DC(\flowprof,\statet) \text{ and } \sum_{\statet\in\states} \prior(\statet)\flow_{\act}\cost_{\actalt}(\flowprof,\statet),\text{ with } \act,\actalt\in\actions.
 \end{equation*}
Thus,  program \eqref{eq:designer-program} can be expressed via the convex hull of the joint range of those functions. 
The dimension of this set gives the upper bound on the cardinality of the support.

\begin{lemma}
\label{pr:optimal-BCWE}
Program \eqref{eq:designer-program} admits a solution $\outcome$ with finite support, whose cardinality is at most $\abs*{\states}\parens*{\abs*{\cup_{\pop}\parens*{\actions^{\pop}\times\actions^{\pop}}}+1}$.
\end{lemma}

\begin{proof}
First,  we show that a \ac{BCWE} $\outcome \in \simplex(\flows)^\states$ can equivalently be represented by some $\tilde\outcome\in\simplex(\flows^{\states})$.
To achieve this, Kuhn's Theorem on the equivalence between mixed and behavior strategies is applied to the one-player game tree where $\statet$ is observed before choosing $\flow$.
Notice that the set $\flows$ is infinite; to cover this setting, we invoke  \citet[theorem~II.1.6, page 63]{MerSorZam:CUP2015}.
To be more explicit, if we define
\begin{equation}
\label{eq:flows-to-states}
\flows^{\states} \coloneqq \braces*{(\flowprof(\statet))_{\statet\in\states} \colon \flowprof(\statet)\in\flows},
\end{equation}
the mediator draws  a vector  $\tilde\flowprof\in\flows^{\states}$ according to the product distribution   $\tilde\outcome$ over $\flows^{\states}$, defined as follows: for any family of Borel sets $\borelset_{\statet} \in \borel(\flows)$, $\statet\in\states$,
\begin{equation}
\label{eq:mu-tilde-def}
\tilde\outcome\parens*{\times_{\statet\in\states}\borelset_{\statet}}
\coloneqq
\prod_{\statet\in\states}\outcome(\borelset_{\statet}\mid\statet).
\end{equation}

Program~\eqref{eq:designer-program} can be written as
\begin{equation*}
\begin{split}
&\min_{\tilde\outcome \in \simplex(\flows^{\states})} \int\sum_{\statet} \prior(\statet)  \DC(\flowprof(\statet),\statet)\diff\tilde\outcome(\tilde\flowprof),\\
&\text{ s.t. }\int\sum_{\statet} \prior(\statet) \flow_{\act}^{\pop}(\statet)\cost_{\act}^{\pop}(\flowprof(\statet),\statet) \diff\tilde\outcome(\tilde\flowprof)
\leq  \int\sum_{\statet} \prior(\statet) \flow_{\act}^{\pop}(\statet)\cost_{\actalt}^{\pop}(\flowprof(\statet),\statet)\diff\tilde\outcome(\tilde\flowprof), 
\quad\forall\pop\in\pops, \forall \act,\actalt\in\actions^{\pop} .
\end{split}
\end{equation*}
For $\pop\in\pops$, $\act,\actalt\in\actions^{\pop}$,
and $f \colon \flows \times \states \to \reals$, define
\begin{align}
\tilde\flowz_{\act,\actalt}^{\pop}(\tilde\flowprof)
&\coloneqq
\sum_{\statet} \prior(\statet) \flow_{\act}^{\pop}(\statet)\cost_{\actalt}^{\pop}(\flowprof(\statet),\statet),\\
\tilde\flowprofz(\tilde\flowprof)
&\coloneqq (\tilde\flowz_{\act,\actalt}^{\pop}(\tilde\flowprof))_{\pop\in\pops,(\act,\actalt)\in \actions^{\pop}\times\actions^{\pop}}
\in \reals^{\cup_{\pop}(\actions^{\pop}\times\actions^{\pop})},\\
\overline{\DC}(\tilde\flowprof)
& \coloneqq \sum_{\statet} \prior(\statet)  \DC(\flowprof(\statet),\statet),\\
\setzc
& \coloneqq \braces*{(\flowprofz, \cost)\in\reals^{\cup_{\pop}(\actions^{\pop}\times\actions^{\pop})}\times\reals : \exists \tilde\flowprof\in \flows^{\states} \text{ s.t. } (\flowprofz, \cost)=(\tilde\flowprofz(\tilde\flowprof), \overline{\DC}(\tilde\flowprof))}.
\end{align}

The objective function and the constraints are linear with respect to $(\tilde\flowprofz(\tilde\flowprof), \overline{\DC}(\tilde\flowprof))$, so Program~\eqref{eq:designer-program} is equivalent to
\begin{equation*}
\min\braces*{\cost \text{ s.t. } (\flowprofz, \cost)\in\co(\setzc) \text{ and }  \flowz^{\pop}_{\act,\act}\leq\flowz_{\act,\actalt}^{\pop},\ \forall\pop\in\pops, \forall \act,\actalt\in \actions^{\pop}},
\end{equation*}
where $\co(\setzc)$ is the convex hull of $\setzc$.
Because $\setzc$ is connected, by Caratheodory's theorem, any point in $\co(\setzc)$ can be obtained as a convex combination of $\abs*{\cup_{\pop}(\actions^{\pop}\times\actions^{\pop})}+1$ points \citep{Fen:MA1929}. 
The value of the objective function at any \ac{BCWE} can be obtained with  a \ac{BCWE} that randomizes over a set of flows with cardinality at most $\abs*{\states}\parens*{\abs*{\cup_{\pop}\parens*{\actions^{\pop}\times\actions^{\pop}}}+1}$, because any $\tilde\flowprof$  induces at most $\abs{\states}$ different flows.  
\end{proof}

\cref{pr:optimal-BCWE} has the following implication.
\begin{lemma}

\label{le:dense}
The set of \acp{BCWE} with finite support is dense in the set of \acp{BCWE}.
\end{lemma}

\begin{proof}
Call $\closure$ the closure of the set of \acp{BCWE} with finite support and assume, \emph{ad absurdum}, the existence of a \ac{BCWE} $\outcome$ outside $\closure$. 
Because the set $\closure\subseteq (\simplex(\flows))^\states$ is convex and closed, from the separation theorem, for each $\statet$, there exists a continuous function $f(\argdot,\statet):\flows\to \reals$ such that
\begin{equation}
\label{eq:separation}
\sum_{\statet} \prior(\statet)\int f(\flowprof,\statet)\diff\outcome(\flowprof\mid\statet)<\inf_{\outcomealt\in \closure}\braces*{ \sum_{\statet} \prior(\statet)\int f(\flowprof,\statet)\diff\outcomealt(\flowprof\mid\statet)},
\end{equation}
which contradicts \cref{pr:optimal-BCWE}.
\end{proof}

\begin{proof}[Proof of \cref{pr:convergence-converse}] 
We prove the result for finite games with $\abs*{\pops}\nplayers$ players with $\nplayers$ players per population, where each player in population $\pop$ has weight $\sizepop^{\pop}/\nplayers$.

\noindent
\ref{it:pr:convergence-converse-a}
From \cref{le:dense} it is enough to prove the result for a \ac{BCWE} $\outcome$  with finite support. 
Dividing by $\sizepop^{\pop}$, the \ac{BCWE} condition is equivalent to
\begin{equation}
\label{eq:BCWEconversebis}
\forall\pop\in\pops,\forall \act,\actalt\in\actions^{\pop},\ \sum_{\statet,\flowprof}\prior(\statet)\outcome(\flowprof\mid\statet)\frac{\flow_{\act}^{\pop}}{\sizepop^{\pop}}\cost_{\act}^{\pop}(\flowprof,\statet)
\leq \sum_{\statet,\flowprof}\prior(\statet)\outcome(\flowprof\mid\statet)\frac{\flow_{\act}^{\pop}}{\sizepop^{\pop}} \cost_{\actalt}^{\pop}(\flowprof,\statet).
\end{equation}

Let $\flows^{*} \coloneqq \cup_{\statet} \supp\outcome(\,\cdot\mid\statet)$. 
We approximate  the numbers $\flow_{\act}^{\pop}/\sizepop^{\pop}$ with sequences of rationals for  $\pop\in\pops$, $\act\in \actions^{\pop}$, and $ \flowprof\in \flows^{*}$.
For every integer $\nplayers$, there exist integers 
$(\nappr_{\act}^{\pop,\nplayers}(\flowprof))_{\pop\in\pops,\act\in\actions^{\pop}, \flowprof\in \flows^{*}}$ such that, for all $\flowprof\in \flows^{*}$, $\pop\in\pops$, and $\act\in \actions^{\pop}$, 
\begin{equation}
\label{eq:approx}
\sum_{\act\in\actions^{\pop}} \nappr_{\act}^{\pop,\nplayers}(\flowprof)=\nplayers \quad \text{and}\quad \abs*{\frac{\nappr_{\act}^{\pop,\nplayers}(\flowprof)}{\nplayers}-\frac{\flow_{\act}^{\pop}}{\sizepop^{\pop}}}\leq  \error_{\nplayers},
\end{equation}
with $\lim_{\nplayers\to\infty}\error_{\nplayers} = 0$. 
The flow profile 
\begin{equation}
\label{eq:approx-flow}
\parens*{\sizepop^{\pop}\frac{\nappr_{\act}^{\pop,\nplayers}(\flowprof)}{\nplayers}}_{\pop\in\pops,\act\in\actions^{\pop}}
\end{equation}
is denoted by $\flowprof_{\nplayers}$.

We construct $\bce^\nplayers:\states\to \simplex\parens*{\bigtimes_{\pop} (\actions^{\pop})^{\nplayers^{\pop}}}$ as follows: 
Conditionally on state $\statet$, the  mediator draws $\flowprof\in \flows^{*}$ with probability $\outcome(\flowprof\mid\statet)$, then, for each population $\pop$, recommends action $\act$ to a subset of players of cardinality $\nappr_{\act}^{\pop,\nplayers}(\flowprof)$, chosen uniformly from population $\pop$. 
Conditionally on $(\statet,\flowprof)$, the probability that player $\play$ in population $\pop$ is recommended $\act$ is $\nappr_{\act}^{\pop,\nplayers}(\flowprof)/\nplayers$. 
The total probability  that player $\play$ in population $\pop$ is recommended $\act$ is 
\begin{equation}
\label{eq:prob-chosen}
\Prob^{\pop,\nplayers}(\act) \coloneqq \sum_{\statet,\flowprof}\prior(\statet)\outcome(\flowprof\mid\statet)\frac{\nappr_{\act}^{\pop,\nplayers}(\flowprof)}{\nplayers}.
\end{equation}
Notice that 
\begin{equation}
\label{eq:prob-k-a}
\Prob^{\pop,\nplayers}(\act)\xrightarrow[\nplayers\to\infty]{} \sum_{\statet,\flowprof}\prior(\statet)\outcome(\flowprof\mid\statet)\frac{\flow_{\act}^{\pop}}{\sizepop^{\pop}} =: \Prob^{\pop}(\act).
\end{equation} 
If $\Prob^{\pop}(\act)=0$, then $\Prob^{\pop,\nplayers}(\act)$ is arbitrarily small for large $\nplayers$.
Hence, deviating after being recommended $\act$ cannot bring a profit that is larger than $\varepsilon$ for such  $\nplayers$.  Conditionally on being recommended $\act$, the expected cost that player $\play$ in population $\pop$ incurs, when playing  $\actalt$,  is 
\begin{equation}
\label{eq:exp-cost-a-b}
\frac{1}{\Prob^{\pop,\nplayers}(\act)} \sum_{\statet,\flowprof}\prior(\statet)\outcome(\flowprof\mid\statet)\frac{\nappr_{\act}^{\pop,\nplayers}(\flowprof)}{\nplayers}\cost_{\actalt}^{\pop}\parens*{\flowprof_{\nplayers}^{-\pop},\flowprof^{\pop}_{\nplayers}+\frac{\sizepop^{\pop}}{\nplayers}(\dirac^{\pop}_{\actalt}-\dirac_{\act}^{\pop}),\statet}.
\end{equation}
The triangular inequality gives
\begin{equation}
\label{eq:ineq-1}
\begin{split}
&\abs*{\frac{\nappr_{\act}^{\pop,\nplayers}(\flowprof)}{\nplayers}\cost_{\actalt}^{\pop}\parens*{\flowprof_{\nplayers}^{-\pop},\flowprof^{\pop}_{\nplayers}+\frac{\sizepop^{\pop}}{\nplayers}(\dirac^{\pop}_{\actalt}-\dirac_{\act}^{\pop}),\statet}
-\frac{\flow_{\act}^{\pop}}{\sizepop^{\pop}} \cost_{\actalt}^{\pop}(\flowprof)} \\
&\leq
\frac{\nappr_{\act}^{\pop,\nplayers}(\flowprof)}{\nplayers}
\abs*{\cost_{\actalt}^{\pop}
\parens*{\flowprof_{\nplayers}^{-\pop},\flowprof^{\pop}_{\nplayers}+\frac{\sizepop^{\pop}}{\nplayers}(\dirac^{\pop}_{\actalt}-\dirac_{\act}^{\pop}),\statet}
-\cost_{\actalt}^{\pop}(\flowprof_{\nplayers},\statet)}
+
\abs*{\frac{\nappr_{\act}^{\pop,\nplayers}(\flowprof)}{\nplayers}\cost_{\actalt}^{\pop}(\flowprof_{\nplayers},\statet)-\frac{\flow_{\act}^{\pop}}{\sizepop^{\pop}} \cost_{\actalt}^{\pop}(\flowprof,\statet)}.
\end{split}
\end{equation}
Thus,
\begin{equation}
\label{eq:ineq-2}
\abs*{\frac{\nappr_{\act}^{\pop,\nplayers}(\flowprof)}{\nplayers}\cost_{\actalt}^{\pop}\parens*{\flowprof_{\nplayers}^{-\pop},\flowprof^{\pop}_{\nplayers}+\frac{\sizepop^{\pop}}{\nplayers}(\dirac^{\pop}_{\actalt}-\dirac_{\act}^{\pop}),\statet}
-\frac{\flow_{\act}^{\pop}}{\sizepop^{\pop}} \cost_{\actalt}^{\pop}(\flowprof,\statet)}
\leq
\modcont\parens*{\frac{\sizepop}{\nplayers}}+\modcont(\error_{\nplayers})=:\frac{\varepsilon_{\nplayers}}{2},
\end{equation}
where $\modcont(\argdot)$ is a modulus of continuity ($\lim_{\error\searrow 0}\modcont(\error)=0$) common to all mappings $\flowprof\mapsto \cost_{\act}^{\pop}(\flowprof,\statet)$, $\flowprof\mapsto \flow_{\act}^{\pop} \cost_{\actalt}^{\pop}(\flowprof,\statet)$,  for $\pop\in\pops$ and $(\act,\actalt)\in \actions^{\pop}\times\actions^{\pop}$. 
This modulus of continuity exists because all these mappings are uniformly continuous on the compact $\flows$.
 It follows from \cref{eq:BCWEconversebis} that no unilateral deviation can lead to a profit larger that $\varepsilon_{\nplayers}$.

\noindent
\ref{it:pr:convergence-converse-b}
 Take a \ac{BDWE}  $\flowprof(\argdot)$, fix a number of players $\nplayers$ for each population $\pop$ and construct $\bce^{\nplayers}$ as follows. 
For each state $\statet$ and each population $\pop$, each player is recommended action $\act$ with probability $\flow_{\act}^{\pop}(\statet)/\sizepop^{\pop}$. Recommendations are \iid across players in each population and independent across populations. 
Let $\recc_{\act}^{\pop}(\nplayers)=(\sizepop^{\pop}/\nplayers)\sum_{\play=1}^{\nplayers}\ind\{\act_{\play}^{\pop}=\act\}$ denote the weighted average number of players in population $\pop$ who are recommended action $\act$ and let $\reccprof^{\pop}(\nplayers) \coloneqq(\recc_{\act}^{\pop}(\nplayers))_{\act\in\actions^{\pop}}$.
Under  $\bce^{\nplayers}$, a player $\play$ in population $\pop$ who plays  $\actalt$, conditionally on being  recommended $\act$, incurs the expected cost $\Cost_{\act,\actalt}^{\pop}(\bce^{\nplayers})$, where
\begin{equation}
\label{eq:Cost-i-a-b}
\begin{split}
\Cost_{\act,\actalt}^{\pop}(\bce^{\nplayers})
&=\frac{1}{\Prob^{\pop}(\act)}\sum_{\statet} \prior(\statet)
\Expect^{\nplayers}\bracks*{\ind\{\act_{\play}^{\pop}=\act\}\cost_{\actalt}^{\pop}
\parens*{\reccprof^{-\pop}(\nplayers),\recc^{\pop}(\nplayers)+(\sizepop^{\pop}/\nplayers)(\dirac^{\pop}_{\actalt}-\dirac_{\act}^{\pop}),\statet}},\\
&=\frac{1}{\Prob^{\pop}(\act)}\sum_{\statet} \prior(\statet)\frac{\flow_{\act}^{\pop}(\statet)}{\sizepop^{\pop}}
\Expect^{\nplayers}\bracks*{\cost_{\actalt}^{\pop}
\parens*{\reccprof^{-\pop}(\nplayers),\recc^{\pop}(\nplayers)+(\sizepop^{\pop}/\nplayers)(\dirac^{\pop}_{\actalt}-\dirac_{\act}^{\pop}),\statet} \mid  \statet,\act_{\play}^{\pop}=\act},
\end{split}
\end{equation}
with 
\begin{equation}
\label{eq:P-a-k}
\Prob^{\pop}(\act) \coloneqq \sum_{\statet} \prior(\statet)\frac{\flow_{\act}^{\pop}(\statet)}{\sizepop^{\pop}}.
\end{equation}
The product $\Prob^{\pop}(\act)\Cost_{\act,\actalt}^{\pop}(\bce^{\nplayers})$ is the expectation
$\Expect\bracks*{\ind\{\act_{\play}^{\pop}=\act\}(\text{cost of playing } \actalt)}$.
By the law of large numbers,  this quantity converges to
\begin{equation}
\label{eq:lln}
\sum \prior(\statet)\frac{\flow_{\act}^{\pop}(\statet)}{\sizepop^{\pop}}\cost_{\actalt}^{\pop}(\flowprof^{-\pop}(\statet),\flow^{\pop}(\statet),\statet).
\end{equation}
The set of triples $(\pop,\act,\actalt)$ is finite; hence, for any $\varepsilon>0$, there exists $\bar{\nplayers}$ such that for all $\nplayers \geq \bar{\nplayers}$, for all $\pop\in\pops$, and all $\act,\actalt\in\actions^{\pop}$, we have
\begin{equation}
\label{eq:diff<=e}
\abs*{\Prob^{\pop}(\act)\Cost_{\act,\actalt}^{\pop}(\bce^{\nplayers})-\sum \prior(\statet)\frac{\flow_{\act}^{\pop}(\statet)}{\sizepop^{\pop}}\cost_{\actalt}^{\pop}(\flowprof^{-\pop}(\statet),\flow^{\pop}(\statet),\statet)}
\leq\frac{\varepsilon}{2}.
\end{equation} 
Because $\flowprof(\argdot)$ is a  \ac{BDWE}, from \cref{eq:BCWEconversebis} we get  that $\bce^{\nplayers}$ is an $\varepsilon$-\ac{BCE}.
\end{proof}


\section{Correlated equilibria and Wardrop equilibria in potential games}
\label{se:complete}

This section studies \ac{BCWE} in the class of convex potential games with complete information. 
A Bayesian nonatomic game $\game$ with $|\states| = 1$ is simply called a nonatomic game and denoted by $\game = \parens*{\pops, \sizepopprof, \actions,\costprof}$. In this section we restrict attention to this class of games with complete information and remove any reference to the state $\statet$ in the notations. Recall that a \acfi{WE}\acused{WE} of the \acl{ANG}  $\game$ is a flow $\flowprof=(\flowprof^{\pop})_{\pop\in \pops}$  such that for all $\pop\in\pops$ and all $\act, \actalt \in \actions^{\pop}$, we have
 \begin{equation}\label{eq:WEmutltipop} 
\flow_{\act}^{\pop} \cost_{\act}^{\pop}(\flowprof)
\leq  \flow_{\act}^{\pop} \cost_{\actalt}^{\pop}(\flowprof).
\end{equation}

\begin{definition}
\label{de:CWE}
A \acfi{CWE}\acused{CWE} of the \acl{ANG}  $\game$ is a distribution $\outcome\in\simplex(\flows)$ over flows such that for all $\pop\in\pops$ and all $\act,\actalt \in \actions^{\pop}$, we have
 \begin{equation}\label{eq:CWEmutltipop} 
\int \flow_{\act}^{\pop} \cost_{\act}^{\pop}(\flowprof)\diff\outcome(\flowprof)
\leq \int \flow_{\act}^{\pop} \cost_{\actalt}^{\pop}(\flowprof)\diff\outcome(\flowprof).
\end{equation}

A \acfi{CCWE}\acused{CCWE} of the \acl{ANG}  $\game$ is a distribution $\outcome\in\simplex(\flows)$ over flows such that for all $\pop\in\pops$, and all $\actalt \in \actions^{\pop}$, we have
\begin{equation}
\label{eq:CCWE}
\int \sum_{\act\in \actions^{\pop}}\flow_{\act}^\pop \cost_{\act}^\pop(\flowprof) \diff \outcome(\flowprof)
\le
\int \sizepop^\pop \cost_{\actalt}^{\pop}(\flowprof) \diff \outcome(\flowprof).
\end{equation}
The symbols $\WE(\game)$, $\CWE(\game)$  and $\CCWE(\game)$ denote the sets of \ac{WE}, \ac{CWE} and \ac{CCWE} of $\game$, respectively.
\end{definition}

 Remark that \ac{CWE} is simply a \ac{BCWE} under complete information.
 Also, a \ac{CWE} that assigns probability one to a single flow induces no strategic uncertainty and corresponds to a \ac{WE}. 
A \ac{CCWE} is the adaptation of the concept of coarse correlated equilibrium \citep{MouVia:IJGT1978} to nonatomic games: the mediator recommends actions in such a way that obedience yields an ex-ante expected cost which is less or equal to the cost of playing $\actalt$ unconditionally on the recommendation.

It is immediate to see that 
\begin{equation}
\label{eq:implicationCE-simplex}
\simplex(\WE(\game)) \subseteq \CWE(\game) \subseteq \CCWE(\game).
\end{equation} 
However, in general, the inclusions can be strict. The next example, taken from  \citet{MitSaaSai:SAGT2013}, shows that $\simplex(\WE(\game))$ can be a proper subset of $\CWE(\game)$.

\begin{example}[El Farol tapas bar \citep{Art:AER1994}]
\label{ex-El-Farol}
In this \acl{ANG}, there is a single population and the action set $\actions$ is $\braces*{\act,\actalt}$, which represent staying home and going to the bar, respectively. 
The cost functions are:
\begin{equation}
\label{eq:cost-el-farol}
\cost_{\act}(\flowprof) = 1, \quad 
\cost_{\actalt}(\flowprof) = \max\braces*{2-4\flow_{\actalt}, 4\flow_{\actalt}-2}.
\end{equation}
The idea is that it is nice to go to the bar when there is some crowd, but not when the crowd is either too large or too small.
The game $\game$ admits three \acp{WE}: $(1,0)$, $(3/4,1/4)$, and $(1/4,3/4)$, and the total cost is equal to $1$ for all of them.

Consider now the flows $\flowprofz = (1,0)$ and $\flowprofw=(1/2,1/2)$.
The distribution $\outcome$ such that 
\begin{equation}
\label{eq:CWE-el-farol}
\outcome(\flowprofz)=\frac{1}{3},\quad
\outcome(\flowprofw)=\frac{2}{3},
\end{equation}
is a \ac{CWE}, because it satisfies the equilibrium constraints
\begin{align*}
\frac{2}{3} =
\frac{1}{3} \flowz_{\act} \cost_{\act}(\flowprofz) +
\frac{2}{3} \floww_{\act} \cost_{\act}(\flowprofw) 
&\le
\frac{1}{3} \flowz_{\act} \cost_{\actalt}(\flowprofz) +
\frac{2}{3} \floww_{\act} \cost_{\actalt}(\flowprofw) =
\frac{2}{3},\\
0 =
\frac{1}{3} \flowz_{\actalt} \cost_{\actalt}(\flowprofz) +
\frac{2}{3} \floww_{\actalt} \cost_{\actalt}(\flowprofw) 
&\le
\frac{1}{3} \flowz_{\actalt} \cost_{\act}(\flowprofz) +
\frac{2}{3} \floww_{\actalt} \cost_{\act}(\flowprofw) =
\frac{1}{3}.
\end{align*}
This \ac{CWE} is not a mixture of \acp{WE}.  
Therefore, $\simplex(\WE(\game)) \subsetneq \CWE(\game)$. 
Moreover, the total cost of this \ac{CWE} is $2/3$, which is  lower than the total cost of each \ac{WE}.
\end{example}

Now we  introduce an important class of \aclp{ANG} for which equality holds in \eqref{eq:implicationCE-simplex}.

\begin{definition}
\label{de:potential}
A \acl{ANG}  is a \emph{potential game} if there exists an open neighborhood  $\widehat{\flows}$ of $\flows$ and a continuously differentiable function  $\potential \colon \widehat{\flows} \to \reals$  such that, for every $\pop\in \pops$, $\act\in\actions$ and $\flowprof\in\flows$, we have
\begin{equation}
\label{eq:potential}
\frac{\partial\potential(\flowprof)}{\partial\flow^{\pop}_{\act}} = \cost^{\pop}_{\act}(\flowprof).
\end{equation}
The function $\potential$ is called the \emph{potential} of $\game$. 
\end{definition}

Any congestion game is a  potential game. 
Consider a congestion game with resource set $\edges$, action set  $\actions^\pop$ for  population $\pop$, and continuous nondecreasing cost function $\cost_{\edge} \colon \reals_{+} \to \reals_{+}$ for resource $\edge$. 
This game admits the potential function
\begin{equation}
\label{eq:potential-congestion}
\potential(\flowprof) =\sum_{\edge\in\edges} \int_{0}^{\load_{\edge}} \cost_{\edge}(u) \diff u,
\end{equation}
where $\load_{\edge} = \sum_\pop \sum_{\act\in \actions^\pop, \act\ni\edge} \flow_{\act}^{\pop}$ is the load on resource $\edge$ induced by the flow $\flowprof$. 
If cost functions are nondecreasing, this function is convex.
All  minimizers of this potential function with nondecreasing costs have the same cost profiles $(\cost_\edge(\load_\edge))_{\edge\in\edges}$.  
In other words, although \acp{WE} of a congestion game with nondecreasing costs are not necessarily unique, they  all have  the same costs.

Note that the El Farol game (\cref{ex-El-Farol}) has a potential given by
\begin{equation}
\label{eq:potential-farol}
\potential(1-\flow_{\actalt},\flow_{\actalt}) =
\begin{cases}
1 + \flow_{\actalt} - 2 \flow_{\actalt}^{2} & \text{ if $\flow_{\actalt} \leq \frac{1}{2}$}, \\
2 -3  \flow_{\actalt} + 2 \flow_{\actalt}^{2} & \text{ if $\flow_{\actalt} \geq \frac{1}{2}$,} 
\end{cases}
\end{equation}
which is not convex. 

\vskip0.2cm
The next proposition shows that, for \aclp{ANG} with a convex potential, distributions over \acp{WE} exhaust the set of \acp{CCWE} (and therefore \acp{CWE}),  and equilibrium costs are unique.

\begin{proposition}
\label{pr:potential-convex-hull-equilibria}
If a \acl{ANG}  $\game$ has a convex potential, then 
\begin{equation}
\label{eq:equilibrium-sets-CWE}
\CCWE(\game) = \CWE(\game)=\simplex(\WE(\game)).
\end{equation}
In addition, all \ac{WE} (and therefore all \ac{CCWE} and \ac{CWE}) have the same costs profiles: all actions with  positive flow have the same cost in all equilibria.
\end{proposition}

We first recall the following lemma.
\begin{lemma}
\label{le:WE-min-Phi}
If a nonatomic game has a convex potential, then the set of \acp{WE} is the set of minimizers of $\potential$.
\end{lemma} 

\begin{proof} This result follows from \citet[proposition~3.1]{San:JET2001}, who proves that the set of \acp{WE} is the set of \ac{KKT} points of the minimization problem $\min\{\potential(\flowprof) \colon \flowprof\in\flows\}$.
The \ac{KKT} points are the minimizers of $\potential$, because  $\potential$ is convex.
\end{proof}

\begin{proof}[Proof of \cref{pr:potential-convex-hull-equilibria}]
From \cref{le:WE-min-Phi} we have  
\begin{equation}
\label{eq:simplex-potential}
\simplex(\argmin\potential) = \argmin_{\outcome\in\simplex(\flows)}\Expect_{\outcome}\bracks*{\potential} \subseteq \CWE(\game).
\end{equation} 
Conversely, take a \ac{CCWE} $\outcome$  and suppose that there exists $\flowprofz\in \flows$ such that $\potential(\flowprofz)<\int\potential(\flowprof)\diff\outcome(\flowprof)$. 
Using the fact that $\potential(\flowprofz)-\potential(\flowprof)\ge\nabla\potential(\flowprof)(\flowprofz-\flowprof)$ and integrating, we obtain
\begin{equation*}
\int \nabla \potential(\flowprof)(\flowprofz-\flowprof)\diff\outcome(\flowprof)
\leq \potential(\flowprofz)-\int\potential(\flowprof)\diff\outcome(\flowprof)<0.
\end{equation*}
We have 
\begin{equation*}
\nabla \potential(\flowprof)(\flowprofz-\flowprof)=\sum_{\pop\in\pops}\sum_{\act\in \actions^{\pop}}(\flowz_{\act}^{\pop}-\flow_{\act}^{\pop})\cost_{\act}^{\pop}(\flowprof),
\end{equation*}
so
\begin{equation*}
\sum_{\pop\in\pops}\sum_{\act\in\actions^{\pop}}\int \flow_{\act}^{\pop}\cost_{\act}^{\pop}(\flowprof)\diff\outcome(\flowprof)
>\sum_{\pop\in\pops}\sum_{\act\in\actions^{\pop}}\flowz_{\act}^{\pop}\int \cost_{\act}^{\pop}(\flowprof)\diff\outcome(\flowprof)
\geq\sum_{\pop\in\pops} \sizepop^\pop \min_{\act\in \actions^{\pop}}  \int  \cost_{\act}^{\pop}(\flowprof)\diff\outcome(\flowprof).
\end{equation*}
Thus, there exist $\pop\in\pops$ such that
\begin{equation*}
\sum_{\act\in\actions^{\pop}}\int \flow_{\act}^{\pop}\cost_{\act}^{\pop}(\flowprof)\diff\outcome(\flowprof)
>\sum_{\act\in\actions^{\pop}}\flowz_{\act}^{\pop}\int \cost_{\act}^{\pop}(\flowprof)\diff\outcome(\flowprof)
\geq \sizepop^\pop \min_{\act\in \actions^{\pop}}  \int \cost_{\act}^{\pop}(\flowprof)\diff\outcome(\flowprof),
\end{equation*}
and  $\actalt \in\actions^{\pop}$ such that
\begin{equation*}
\sum_{\act\in\actions^{\pop}}\int \flow_{\act}^{\pop}\cost_{\act}^{\pop}(\flowprof)\diff\outcome(\flowprof)
>\sum_{\act\in\actions^{\pop}}\flowprofz_{\act}^{\pop}\int \cost_{\act}^{\pop}(\flowprof)\diff\outcome(\flowprof)
\geq \sizepop^\pop \int \cost_{\actalt}^{\pop}(\flowprof)\diff\outcome(\flowprof).
\end{equation*}
This contradicts \eqref{eq:CCWE}, so $\outcome$ cannot be a \ac{CCWE}.

Next, we show that all \acp{WE} have the same costs. 
We know from \cref{le:WE-min-Phi} that the \acp{WE} minimize the potential, \ie they are the solutions of the following convex optimization problem:
\begin{equation*}
\min\braces*{\potential(\flowprof) : \forall \pop\in\pops, \sum_{\act\in \actions^\pop} \flow_{\act}^{\pop}=\sizepop^{\pop}, \forall\pop\in\pops, \forall \act\in\actions^\pop, \flow_{\act}^{\pop}\geq 0},
\end{equation*}
whose Lagrangian is 
\begin{equation*}
\lagrange(\flowprof,\multiplprof)=\potential(\flowprof)-\sum_{\pop\in\pops}\multipl^{\pop}\parens*{\sum_{\act\in\actions^\pop} \flow_{\act}^{\pop}}-\sum_{\pop\in\pops}\sum_{\act\in\actions^\pop}\multipl_{\act,\pop}\flow_{\act}^{\pop},
\end{equation*}
with $\multiplprof=
\parens*{\parens*{\multipl^{\pop}}_{\pop\in\pops},(\multipl_{\act,\pop})_{\act\in\actions^\pop, \pop\in\pops})}$. 
From the \ac{KKT} theorem, $\widehat\flowprof$ is a solution if and only if there exists $\widehat\multiplprof$ such that $(\widehat\flowprof,\widehat\multiplprof)$ satisfies the \ac{KKT} conditions:
For all $\pop\in\pops$ and $\act\in\actions^\pop$,
\begin{equation*}
\frac{\partial \potential}{\partial \flow_{\act}^{\pop}}(\widehat\flowprof)=\widehat{\multipl}^{\pop} + \widehat{\multipl}_{\act}^{\pop};
\quad \widehat{\multipl}_{\act}^{\pop}\cdot \widehat{\flow}_{\act}^{\pop} = 0;
\quad \widehat{\multipl}_{\act}^{\pop}\ge 0;
\quad \sum_{\act\in\actions^\pop}\widehat{\flow}_{\act}^{\pop}=\sizepop^\pop.
\end{equation*}
This condition is satisfied if and only if for all $(\flowprof,\multiplprof)$,
\begin{equation*}
\lagrange(\widehat\flowprof,\multiplprof) 
\leq \lagrange(\widehat\flowprof,\widehat\multiplprof)
\leq \lagrange(\flowprof,\widehat\multiplprof),
\end{equation*}
\citep[see][theorem~28.3, page~281]{Roc:PUP1970}.
From the exchange property, if $(\widehat\flowprof,\widehat\multipl)$ and $(\bar\flowprof,\bar\multipl)$ are such saddle points, then $(\widehat\flowprof,\bar\multipl)$ and $(\bar\flowprof,\widehat\multipl)$ are also saddle points.

Consider  the relative interior of the set $\WE(\game)$. 
For each $\pop\in\pops$, there exists a subset of actions $\subactions^{\pop}\subseteq \actions^{\pop}$ such that $\times_{\pop\in\pops}\subactions^{\pop}$ is the support of all flows in this relative interior. 
To see this, notice that for each pair of \acp{WE} $\flowprof,\flowprofz$ such that $\flowprof$ is in the relative interior and for each $t\in(0,1]$, we have that $t\flowprof+(1-t)\flowprofz$ is also in the relative interior. 
Therefore, for each $\act,\pop$, whenever $\flow_{\act}^{\pop}>0$ for some \ac{WE}, this must be true for all points in the relative interior of \ac{WE}. 

Consider now two points $\widehat\flowprof,\bar\flowprof$ in the relative interior of $\WE(\game)$. For every $\act\in \subactions^{\pop}$, we have $\widehat{\flow}_{\act}^{\pop} > 0$, 
$\bar\flow_{\act}^{\pop} > 0$ and
\begin{equation*}
\frac{\partial \potential}{\partial \flow_{\act}^{\pop}}(\widehat\flowprof)=\widehat{\multipl}^{\pop} = \bar\multipl^{\pop}=\frac{\partial \potential}{\partial \flow_{\act}^{\pop}}(\bar\flowprof).
\end{equation*}
Thus, for every $\pop\in\pops$ and $\act\in\subactions^{\pop}$, $\cost_{\act}^{\pop}(\widehat\flowprof)=\cost_{\act}^{\pop}(\bar\flowprof)$, and all points in the relative interior of $\WE(\game)$ have the same costs. 
By continuity, all points in $\WE(\game)$  have the same costs (notice that the support can only shrink when approaching the boundary).  
\end{proof}

\cref{pr:potential-convex-hull-equilibria} implies that mediation does not help as a tool for inducing other equilibrium outcomes than \ac{WE}. \citet[Theorem~1]{DiaMitRusSai:WINE2009} showed that in a particular class of routing games, the smallest total cost achieved by a \ac{CWE} cannot be smaller than the smallest total cost achieved by a \ac{WE}. 
\cref{pr:potential-convex-hull-equilibria} significantly
generalizes the result of \citet{DiaMitRusSai:WINE2009} in two directions. 
First, \cref{pr:potential-convex-hull-equilibria} applies to all games with a convex potential.
Second, it implies that mediation has no value regardless of the welfare objective, not only for minimizing the total cost.

The  first part of the proof of \cref{pr:potential-convex-hull-equilibria},  showing that any \ac{CCWE} is a distribution over \ac{WE}, follows similar steps as the proof of \citet[theorem~1]{Ney:IJGT1997}. 
Neyman's result states that correlated equilibria are distributions of Nash equilibria in potential games with finitely many players, convex sets of actions, and a convex potential. 
There are several differences between   \cref{pr:potential-convex-hull-equilibria} and \citet{Ney:IJGT1997}: finite vs.\ infinite set of players, convex vs.\ finite set of actions.  
Also, the   games in  \cite{Ney:IJGT1997} are potential games in the sense of \citet{MonSha:GEB1996}:
 differences of cost functions along unilateral deviations are equal to differences of the potential. 
 Lastly,  \citet{Ney:IJGT1997} considers only correlated equilibria and not coarse correlated equilibria. 
\citet{MouRaySen:JET2014} provide an example of a game satisfying the conditions of \citet{Ney:IJGT1997} where there is a unique correlated equilibrium and a more efficient coarse correlated equilibrium. 
Hence, in the setup of \citet{Ney:IJGT1997} with a convex potential, coarse correlated equilibria do not reduce to distributions over Nash equilibria.
 
Uniqueness of \ac{WE} costs is a well known   property of congestion games with nondecreasing costs.  
The usual proof uses the fact that costs of edges are functions of one-dimensional variables (the loads) and that the potential is given by the sum of  the integrals of these function  \citep[see][theorem~18.8 and its proof]{Rou:AGT2007}. 
\cref{pr:potential-convex-hull-equilibria} generalizes it to all games with a convex potential. 
In our proof, we use the fact that \acp{WE} are minimizers of the potential and that pairs of optimal solutions and Lagrange multipliers are saddle-points of the Lagrangian function.

We have previously observed that a network routing game with multiple populations has a convex potential when populations only differ by their feasible actions, \ie by their origin-destination pairs and feasible routes  \citep[see \eg][proposition 18.11]{Rou:AGT2007}.
Hence, \cref{pr:potential-convex-hull-equilibria} applies to such games, and  mediation has no value. 
However, if the cost functions differ across populations, then the multi-population routing game does not necessarily have a convex potential.  
The next example shows a   two-population routing game with different cost functions, where there exists a \ac{CWE} that Pareto dominates the unique \ac{WE}.

\begin{example} 
\label{ex:trucks}
Consider the following two-population routing game.
The set of actions is $\actions = \braces*{\act,\actalt}$ for both populations.  
The costs functions in population~1 are given by  $\cost_{\act}^{1}(\flowprof) = 1$ and $\cost_{\actalt}^{1}(\flowprof) = 2\flow_{\actalt}^{1}$. 
The cost functions in population 2 are given by $\cost_{\act}^{2}(\flowprof) = 2$ and 
\begin{equation}
\label{eq:cost-b-2}
\cost_{\actalt}^{2}(\flowprof) = 
\begin{cases}
2 \flow_{\actalt}^{1} & \text{if } \flow_{\actalt}^{1} \leq 1/2 \\
1  & \text{if } \flow_{\actalt}^{1} > 1/2. \\
\end{cases}
\end{equation}
Action $\actalt$ is  strictly dominant for population~2, so the flow of population~2 on action $\actalt$ is $1$ in every \ac{CWE}. 
The unique \ac{WE} is $\flow_{\actalt}^{1} = 1/2$ and $\flow_{\actalt}^{2}=1$. 
Consider now the \ac{CWE} in which the flow of population~1 on action $\actalt$ is equal to $1$ with probability $1/2$ and is equal to $0$ with probability $1/2$. 
The total cost of population~1 is unchanged, but the  expected  total cost of population~2 is strictly lower than in the \ac{WE}: 
\begin{equation*}
\frac12 \cost_{\actalt}^{2}(1) + \frac12 \cost_{\actalt}^{2}(0) = 1/2 < \cost_{\actalt}^{2}(1/2) = 1.
\end{equation*}

\end{example}

Combining \cref{pr:convergence-A-NA} (applied to games with complete information) and \cref{pr:potential-convex-hull-equilibria}, we get the following  important corollary:

\begin{corollary}
\label{co:convergence-potential}
Let  $\game$ be a  \acl{ANG}  with convex potential. Consider a sequence of weights $\braces*{\weight(\nplayers)}$ such that $\max_{\pop,\play}\weight_{\play}^{\pop}(\nplayers)\to_\nplayers 0$, and let $\outcome^{\nplayers}$ be a correlated equilibrium outcome of $\game^{\nplayers}$ for each integer $\nplayers$. 
Then any weak$^{*}$ accumulation point $\outcome$ of the sequence $\braces*{\outcome^{\nplayers}}$ belongs to $\simplex(\WE(\game))$.
\end{corollary}

This corollary shows that the value of mediation, that is, the additional welfare due to correlation,  tends to zero when the weight of each player tends to zero, \ie when the impact of each player's action  on other players' costs becomes negligible. 
\citet{AshMonTen:JAIR2008} showed that correlated equilibria may significantly decrease the \acl{TC} over Nash equilibria (the value of mediation is  positive) in  $\nplayers$-player routing games, even for a large number of players. 
However, in their examples, the weights of some  players do not tend to zero. 
Also,  our model takes  as primitives continuous cost functions defined for all real valued flows.  In the  $\nplayers$-player games, costs are given by those continuous functions evaluated at the flows induced by the actions of the  $\nplayers$ players, for any $\nplayers$. By contrast, in the examples of \citet{AshMonTen:JAIR2008}, cost functions are tuned to the number of players.

Given that the concepts of \ac{BDWE} and \ac{BCWE} are extensions of---respectively---\ac{WE} and \ac{CWE} to \aclp{BANG}, it is natural to ask whether \cref{pr:potential-convex-hull-equilibria} extends to \aclp{BANG} that admit a convex potential in each state. 
The next example shows that this it not the case in general. 
This is a  single-population routing game with two actions, two states, and no cost externality. 
We find that there is a flow profile in the support of a \ac{BCWE} which is not a \ac{BDWE} flow profile. 
Hence, even if the game admits a convex potential in each state, distributions over \acp{BDWE} do not exhaust the set of \acp{BCWE}.

\begin{example}\label{ex-BCWEnotBDWE}
Consider a \acl{BANG} with a single population, where the  action set $\actions$ is $\braces*{\act,\actalt}$, the state set $\states$ is $\braces*{0,1}$, and the prior is uniform. 
The cost functions are given by $\cost_{a}(\flowprof,\statet)=\statet$ and $\cost_{b}(\flowprof,\statet)=1/3$ for every $\statet$ and $\flowprof$. It is easy to verify that for any  $\alpha \in [0,\frac12]$, there is a \ac{BCWE} such in state $\statet = 0$ the flow is $\flowprof(\statet) = (1,0)$ with probability 1, and in state $\statet = 1$ the flow is $\flowprof(\statet) = (1,0)$ with probability $\alpha$ and $\flowprof(\statet) = (0,1)$ with probability $1 - \alpha$. However, there is no \ac{BDWE} in which the flow is $\flowprof(\statet) = (1,0)$ in state $\statet =1$.
\end{example}


\section{No-regret  and \aclp{CCWE}}
\label{se:no-regret}

The notion of \ac{CCWE} is also related to no-regret procedures. 
Define a no-regret sequence of flows as follows (see \eg \citealp{BluEveLig:TC2010}).

\begin{definition}
\label{de:no-regret}
A sequence of flows $\flowprof(t)$, $t\in\naturals$, has no regret if for all $\pop\in\pops$ we have
\begin{equation}\label{eq:noregret}
\frac{1}{T}\sum_{t=1}^T\sum_{\act\in \actions^\pop}\flow_{\act}^{\pop}(t)\cost_{\act}^{\pop}(\flowprof^\pop(t))
\leq
\min_\actalt \frac{1}{T}\sum_{t=1}^T \sizepop^\pop \cost_{\actalt}^{\pop}(\flowprof(t))+\varepsilon_T,
\end{equation}
for every $T\in\naturals$, with $\lim_{T\to+\infty}\varepsilon_T=0$.
\end{definition}

The link with \acl{CCWE} is evident by passing to the limit in \cref{eq:noregret}. This gives:
\begin{observation}
\label{ob:no-regret-CCWE}
If a sequence of  flows $\flowprof(t)$ has no regret, then any weak$^{*}$ accumulation point $\outcome$ of $\frac{1}{T} \sum_{t=1}^T \dirac_{\flowprof(t)}$ is a \acl{CCWE}.
\end{observation}

\citet{BluEveLig:TC2010} consider nonatomic congestion games and show that if a sequence of flows has no external regret, then its average cost converges to the \ac{WE} cost. 
Our approach allows us to generalize this result. 
We have the following generalization of Theorem~4.1 in  \citet{BluEveLig:TC2010}:
\begin{corollary}\label{coro-no-regret}
Consider a nonatomic game $\game$ with convex potential. If a sequence of  flows $\flowprof(t)$ has no regret, then any weak$^{*}$ accumulation point of the sequence $\frac{1}{T}\sum_{t=1}^T\delta_{\flowprof(t)}$  belongs to $\simplex(\WE(\game))$.
\end{corollary}

In particular, in a nonatomic game with convex potential, if a sequence of  flows $\flowprof(t)$ has no regret, then costs converge to the \ac{WE} costs. Another direct consequence is that any accumulation point of the sequence $\frac{1}{T}\sum_{t=1}^T\flowprof(t)$  belongs to $\WE(\game)$.


\subsection*{Acknowledgments}
This work was partially supported by COST Action 16228 GAMENET.
Frederic Koessler acknowledges the support of the ANR (StratCom ANR-19-CE26-0010-01).
Marco Scarsini is a member of GNAMPA-INdAM. 
He acknowledges the support of the GNAMPA project CUP\_E53C22001930001 ``Limiting behavior of stochastic dynamics in the Schelling segregation model'' and the Italian MIUR PRIN 2017 Project ALGADIMAR Algorithms, Games, and Digital Markets.
Tristan Tomala gratefully acknowledges the support of the HEC foundation and ANR/Investissements d'Avenir under grant ANR-11-IDEX-0003/Labex Ecodec/ANR-11-LABX-0047.

\bibliographystyle{apalike}
\bibliography{bibIDLG}

\appendix


\section{List of symbols}
\label{se:symbols}

\begin{longtable}{p{.13\textwidth} p{.82\textwidth}}

$\actions$ &  set of action profiles, defined in \cref{de:BANG}\\
$\actions^{\pop}$ & action set of population $\pop$, defined in \cref{de:BANG}\\
$\act$ & action\\
$\act_{\play}$ & action of player $\play$\\
$\actprof$ & action profile\\
$\actalt$ & action\\
$\subactions^{\pop}$ & subset of $\actions^{\pop}$\\
$\borel$ & Borel $\sigma$-field\\
$\cost_{\act}^{\pop}$ & cost of action $\act$ in population $\pop$, defined in \cref{de:BANG}\\
$\cost_{\edge}$ & cost of resource $\edge$ in a congestion game\\
$\costprof$ & $\parens{\cost_{\act}}_{\act\in\actions}$, defined in \cref{de:BANG}\\
$\Cost_{\play}$ & cost of player $\play$ in the game $\game^{\nplayers}$\\
$\Costprof$ & $\parens*{\Cost_{\play}}_{\play\in\players}$\\
$\Cost_{\act,\actalt}^{\pop}$ & expected cost of a player in population $\pop$ who plays  $\actalt$, conditionally on being  recommended $\act$\\
$\CCWE(\game)$ & \aclp{CCWE} of $\game$\\
$\co(\argdot)$ & convex hull\\
$\CWE(\game)$ & \aclp{CWE} of $\game$\\
$\edge$ & resource in a congestion game\\
$\edges$ & resource set in a congestion game\\
$\Expect^{\nplayers}$ &  expectation corresponding to $\Prob^{\nplayers}$\\
$\pop$ & population\\
$\pops$ & population set\\
$\lagrange$ & Lagrangian\\
$\players$ & player set of the game $\game^{\nplayers}$\\
$\nplayers^{\pop}$ & size of population $\pop$ in the game $\game^{\nplayers}$\\
$\nappr_{\act}^{\pop,\nplayers}$ & number of players in population $\pop$ that are recommended $\act$\\ 
$\prior$ & prior distribution on $\states$, defined in \cref{de:BANG}\\
$\Prob^{\nplayers}$ &  probability measure induced by the prior $\prior$ and the \ac{BCE} $\bce^{\nplayers}$\\
$\Prob^{\pop,\nplayers}(\act)$  & probability  that player $\play$ in population $\pop$ is recommended $\act$, defined in \eqref{eq:P-a-k} \\
$\weight_{\play}^{\pop}$ & weight of player $\play$ in population $\pop$\\
$\WE(\game)$ & \aclp{WE}  of $\game$\\
$\load_{\edge}$ & load on resource $\edge$, defined in \eqref{eq:load-congestion}\\
$\flow_{\act}^{\pop}$ & flow on action $\act$ in population $\pop$\\
$\flowprof$ & flow vector\\
$\eq{\flowprof}$ & equilibrium flow vector\\
$\Flow_{\act}^{\pop}(\nplayers)$ & random flow of population $\pop$ on action $\act$\\
$\flows$ & $\times_{\pop} \flows^{\pop}$, set of feasible flows of the \acl{ANG} $\game$\\
$\flows^{\pop}$ & $\simplex_{\sizepop^{\pop}}(\actions^{\pop})$, set of flows for population $\pop$\\
$\flows^{\states}$ & $\braces*{(\flowprof(\statet))_{\statet\in\states}, \text{ with }\flowprof\in\flows}$, defined in \eqref{eq:flows-to-states}\\
$\recc_{\act}^{\pop}(\nplayers)$ & weighted average number of players in population $\pop$ who are recommended action $\act$\\
$\reccprof^{\pop}(\nplayers)$ & $(\recc_{\act}^{\pop}(\nplayers))_{\act\in\actions^{\pop}}$\\

$\bce^\nplayers$ & \acl{BCE} of the game $\game^{\nplayers}$, defined in \cref{de:BCE}\\
$\sizepop^{\pop}$ & size of population $\pop$, defined in \cref{de:BANG}\\
$\sizepopprof$ & vector of population sizes, defined in \cref{de:BANG}\\
$\game$ & \acl{ANG}\\
$\game^{\nplayers}$ & \acl{AG} with $\nplayers$ players\\
$\dirac_{\act}$ & Dirac measure on $\act$\\
$\simplex(\actions)$ & simplex of probability measures on $\actions$\\

$\simplex_{\sizepop}(\spaceJ)$ & $\braces*{\flowprof\in\reals^\spaceJ \colon \forall \elemj\in \spaceJ, \flow_{\elemj} \geq 0, \sum_{\elemj\in\spaceJ} \flow_{\elemj}  = \sizepop}$, defined in \eqref{eq:simplex-gamma}\\

$\statet$ & state\\
$\states$ & state space, defined in \cref{de:BANG}\\

$\outcome$ & outcome of the game, defined in \cref{de:BANG}\\
$\eq{\outcome}$ & \acl{CE} of the game $\game$\\
$\outcome^{\nplayers}$ & \acl{CE} outcome of the game $\game^{\nplayers}$, defined in \eqref{eq:BCEoutcome}\\
$\prtype$ & mapping from $\states$ to $\simplex(\types)$\\
$\potential$ & potential, defined in \eqref{eq:potential}\\
$\modcont(\argdot)$ & modulus of continuity\\

$\wstarto$ & weak convergence\\

$\ind$ & indicator function\\
\end{longtable}

\end{document}